\documentclass[pdflatex,iicol,sn-apa]{sn-jnl}

\usepackage{graphicx}%

\usepackage{multirow}%
\usepackage{amsmath,amssymb,amsfonts}%
\usepackage{amsthm}%
\usepackage{mathrsfs}%
\usepackage[title]{appendix}%
\usepackage{xcolor}%
\usepackage{textcomp}%
\usepackage{manyfoot}%
\usepackage{booktabs}%
\usepackage{algorithm}%
\usepackage{algorithmicx}%
\usepackage{algpseudocode}%
\usepackage{listings}%

\theoremstyle{thmstyleone}%
\theoremstyle{thmstyletwo}%

\theoremstyle{thmstylethree}%

\begin{document}

\title[Article Title]{Enhancing Black-Litterman Portfolio via Hybrid Forecasting Model Combining Multivariate Decomposition and Noise Reduction}


\author[1]{\fnm{Ziye} \sur{Yang}}\email{mc46468@um.edu.mo}

\author[2]{\fnm{Ke} \sur{Lu}}\email{yc37949@um.edu.mo}

\author[3]{\fnm{Yang} \sur{Wang}}\email{yang.wang1@siat.ac.cn}

\author*[2]{\fnm{Jerome} \sur{Yen}}\email{jeromeyen@um.edu.mo}

\affil[1]{\orgdiv{Institute of Collaborative Innovation}, \orgname{University of Macau}, \orgaddress{\city{Macau}, \postcode{999078}, \country{China}}}

\affil[2]{\orgdiv{Faculty of Technology}, \orgname{University of Macau},  \orgaddress{\city{Macau}, \postcode{999078}, \country{China}}}

\affil[3]{\orgname{Shenzhen Institutes of Advanced Technology}, \orgaddress{\city{Shenzhen}, \postcode{518055}, \country{China}}}

\abstract{Modern portfolio construction demands robust methods for integrating data-driven insights into asset allocation. The Black-Litterman model offers a powerful Bayesian approach to adjust equilibrium returns using investor views to form a posterior expectation along with market priors. Mainstream research mainly generates subjective views through statistical models or machine learning methods, among which hybrid models combined with decomposition algorithms perform well. However, most hybrid models do not pay enough attention to noise, and time series decomposition methods based on single variables make it difficult to fully utilize information between multiple variables. Multivariate decomposition also has problems of low efficiency and poor component quality. In this study, we propose a novel hybrid forecasting model—SSA-MAEMD-TCN— to automate and improve the view generation process. The proposed model combines Singular Spectrum Analysis (SSA) for denoising, Multivariate Aligned Empirical Mode Decomposition (MA-EMD) for frequency-aligned decomposition, and Temporal Convolutional Networks (TCNs) for deep sequence learning to capture complex temporal patterns across multiple financial indicators. Empirical tests on the Nasdaq 100 Index stocks show a significant improvement in forecasting performance compared to baseline models based on MAEMD and MEMD. The optimized portfolio performs well, with annualized returns and Sharpe ratios far exceeding those of the traditional portfolio over a short holding period, even after accounting for transaction costs.}




\maketitle

\section{Introduction}
In 1952, Markowitz introduced Modern Portfolio Theory (MPT) to revolutionize portfolio management by optimizing risk return through the Mean-Variance (MV) framework \citep{markowitzPortfolioSelection1952}. However, MV the framework faces many challenges in practical applications.  Its extreme sensitivity to input parameters often leads to significant changes in asset weights \citep{bestSensitivityMeanVarianceEfficientPortfolios1991}. In addition, traditional MV optimizers amplify the estimation errors and ignore the input uncertainty \citep{michaud1989markowitz}. 

Faced with such limitations, Fisher Black and Robert Litterman proposed the Black-Litterman (BL) model \citep{blackGlobalPortfolioOptimization1992}. The model is based on Bayesian statistics that combine the implied return from market equilibrium with subjective investor views to generate a new posterior distribution of returns, which can be used to build a new portfolio to increase its logic \citep{,heIntuitionBlackLittermanModel2002,satchellDemystificationBlackLitterman2000,idzorek2StepbystepGuide2007}. The most important thing is that investors or fund managers could use such BL framework to justify their extra earning by providing such investor views. 

Despite the advantages of the BL model, how to more accurately predict asset movements to form the investor view is still a significant challenge for researchers. In our review of the relevant literature, we found that most studies use different forecasting models to analyze historical data. These approaches include traditional econometric models for time series that are linearly dominant \citep{beachApplicationBlackLitterman2007a,palombaMultivariateGARCHModels2008,duqiBlackLittermanModel2014a}, as well as hybrid models that incorporate traditional econometric analyzes and machine learning \citep{pyoExploitingLowriskAnomaly2018,karaHybridApproachGenerating2019}.

Artificial intelligence has flourished in recent years. Deep learning methods show significant advantages over statistical methods in capturing complex nonlinear relationships in data through automatic feature engineering and complex neural network structures \citep{gao2020application,TANG2022363}. Decomposition Ensemble models (DEMs), consisting of the combination of deep learning model and the Empirical mode decomposition (EMD) algorithm \citep{huangEmpiricalModeDecomposition1998} or its modifications \citep{wuENSEMBLEEMPIRICALMODE2009,yehComplementaryEnsembleEmpirical2010,torresCompleteEnsembleEmpirical2011a}, produced breakthroughs to the financial market in supporting asset movement forecasting with mixed relationships. Decomposition techniques can effectively capture the different patterns in time series, which can help improve the model's ability to learn linear and nonlinear relationships in the data \citep{zhang2024deep}. DEMs provide better predictive power \citep{shuForecastingStockPrice2020,zhangNovelDeepLearning2020,linForecastingStockIndex2021}. Recently, new approaches based on DEMs have been developed and widely used to support the generation of investor views with better performance of portfolio return \citep{rezaeiIntelligentAssetAllocation2021,baruaUsingFearGreed2023}.

The above decomposition-ensemble methods are based on univariate prediction, while multivariate analyses can better explain stock price movements with more information to include \citep{jiangApplicationsDeepLearning2021,kumbureMachineLearningTechniques2022}. The decomposition algorithm cannot guarantee that the number of IMFs is the same for each variable due to differences in the frequency of data for different variables, preventing multivariate forecasting. Therefore, scholars focused on Multivariate Empirical Mode Decomposition (MEMD) \citep{rehmanMultivariateEmpiricalMode2010}. This method can jointly decompose multidimensional signals and ensure that the IMFs obtained from the decomposition in each dimension are matched in both the number and frequency spectrum. In recent years, Memd-based multivariate decomposition ensemble models (M-DEMs) have been widely used in financial markets and outperformed univariate DEM in terms of prediction accuracy \citep{dengMultistepaheadStockPrice2022,zouForecastingCrudeOil2022,baiIntelligentForecastingModel2023,yaoStockIndexForecasting2023}. Among them, \citet{yaoStockIndexForecasting2023} proposed the MEMD-TCN hybrid model for stock index prediction, and emphasized that the TCN took less memory in training, provided higher accuracy, and avoided the problem of gradient explosion or gradient vanishing, which is common in RNNs.

Research into multivariate decomposition algorithms became popular, with a subset of articles pointing out that MEMD requires longer computation times when dealing with high-dimensional data \citep{langFastMultivariateEmpirical2018,mouradMultivariateGroupsparseMode2023}. In addition to the inefficiency, \cite{caiMAEMDAlignedEmpirical2025} experimentally demonstrated that the decomposition quality of MEMD is far inferior to that of EMD by using periodicity and symmetry metrics. Such limitation affects the subsequent prediction accuracy. Therefore, they proposed a new algorithm - Multivariate Aligned Empirical Modal Decomposition (MA-EMD). Based on EMD, the algorithm measures the difference between IMFs based on the frequency representation of the extreme points that support the alignment with IMFs. So far, no research that applied the MA-EMD algorithm to forecast the movement of financial assets has been found, and this is one of our motivations in exploring its effectiveness and usefulness.

In addition, one issue that cannot be ignored is how to deal with noise in financial data. These noises may originate from the fluctuations of market transactions, interference from external factors, and so on \citep{blackNoise1986}. Many studies have shown that the accuracy of financial time series forecasting can be effectively enhanced by appropriate noise reduction \citep {alrumaihTimeSeriesForecasting2002,hassaniEffectNoiseReduction2010,baoDeepLearningFramework2017,yuHybridModelFinancial2020}. But the majority of the current decomposition-ensemble models do not fully consider the noises that exists in the original data \citep{wangTwoStageDeepEnsemble2024}.  As a result, models might overfit the noisy data during the training process, and the predictions with such models might be biased or underfitting when it is difficult to capture the special patterns contained in the data \citep{aksehirNewDenoisingApproach2024a}.

To address these problems, this paper proposes a new SSA-MAEMD-TCN model based on the "M-DEM" forecasting framework, aiming at improving the forecasting accuracy and removing the bias of the investor view for the Black-Litterman model. The difference between this research and previous research include:  First, we replace MEMD with the MA-EMD algorithm for more efficient and accurate sequence decomposition. Secondly, Singular Spectrum Analysis \citep{broomheadExtractingQualitativeDynamics1986} is very effective in noise reduction of financial data\citep{lahmiriMinuteaheadStockPrice2018,tangPredictionFinancialTime2021,wangDeepLearningIntegrated2024}, so SSA was used to reduce its impact. 

The US stock data from January 2020 to September 2023 was selected to test the proposed model’s predictive performance and the performance of the portfolio constructed. We document the differences in the predictive metrics between SSA-MAEMD-TCN and the benchmark model, as well as compare the performance of the Black-Litterman based on the generated predictive model with the other benchmark portfolio models in terms of return, volatility, and other performance measures over different holding periods.

The contributions of this paper include the following three aspects: 
\begin{itemize}
\item This paper explores the application of the MA-EMD algorithm in financial time series forecasting and proves its superiority in dealing with financial data through empirical research, which provides a reference for subsequent related research.
\item The SSA-MAEMD-TCN model is proposed to improve the approach based on M-DEM: the introduction of MA-EMD supports a more efficient and better-quality decomposition; the introduction of SSA effectively reduces the impacts of noise in financial data on the subsequent forecasting steps.
\item The proposed hybrid model generates more accurate and less biased investor views for the Black-Litterman model. The empirical results show that the methodology generates much higher returns in the U.S. stock market.
\end{itemize}

The article is organized as follows: Section 2 reviews past approaches to generating subjective views and the application of M-DEM and SSA in financial markets. Section 3 delves into the core technique’s theoretical foundations. Section 4 introduces the proposed model and details the research process. Section 5 evaluates the experimental results, focusing on forecast accuracy and portfolio performance. Finally, Section 6 summarizes the main findings and discusses future research directions.

\section{Related Work}
This research focuses on the academic areas of generating investor perspectives to enhance BL portfolio and applying M-DEMs and SSA method in financial market.

Early studies mainly generated subjective views through traditional econometric models. \citet{beachApplicationBlackLitterman2007a} first applied the EGARCH-M model to global asset allocation, dynamically estimated the expected returns and conditional variances of assets in 20 countries, and generated dynamic opinions for the BL model, with returns exceeding the market equilibrium weight and Markowitz optimal allocation. \citet{palombaMultivariateGARCHModels2008} combined the multivariate GARCH model with the BL model and used the FDCC model to estimate and predict asset returns and covariance matrices, achieving better risk-return balance and higher information ratio in tactical asset allocation (TAA). \citet{duqiBlackLittermanModel2014a} used the EGARCH-M model to capture volatility clustering, leverage effects, and asymmetry, construct stock portfolios at different risk levels, and calculate expected excess returns through the BL model, performing well in risk-stratified allocation. 

As machine learning advances, blending econometric and machine learning models effectively generates investor views in BL models. \citet{pyoExploitingLowriskAnomaly2018} combined three machine learning models (Gaussian Process Regression, Support Vector Regression, and Artificial Neural Networks) with the GARCH model to forecast asset volatility in the KS200. They divided assets into high-risk and low-risk groups, constructing a BL portfolio reflecting investor views. Results showed this portfolio had higher Sharpe ratios and positive alpha. \citet{karaHybridApproachGenerating2019} used the GARCH model to predict stock indices and Support Vector Regression to map volatility forecasts to return forecasts, generating investor-view vectors for the BL model. This approach outperformed market indices in the BIST-30 and DJI, showing better portfolio returns and Sharpe ratios across holding periods. \citet{leiBlackLittermanAsset2019} proposed a PCA-AHP framework to tackle overfitting in high-dimensional asset allocation, reducing prediction errors in Chinese stock-market sector indices via dimensionality reduction and weight optimization in the BL model.

Recently, deep learning and decomposition ensemble models (DEMs) have also been gradually applied to generate subjective views and improve the performance of asset portfolios. \citet{rezaeiIntelligentAssetAllocation2021} developed a CEEMD-CNN-LSTM hybrid model for stock price prediction. Investor views based on its predictions significantly enhance asset allocation effectiveness. \citet{baruaDynamicBlackLitterman2022} used a CNN-BiLSTM model to predict future stock prices and a dynamic EWMA covariance matrix to emphasize recent data. Their follow-up study \citep{baruaUsingFearGreed2023} built a CEEMDAN-GRU-XGBoost framework, using GRU to predict sentiment indicators and XGBoost to generate ETF views. The improved BL model outperformed others over six investment periods.

Most DEMs are univariate forecasting models that only decompose and forecast closing prices. Multivariate versions of EMD have been developed, including B-EMD \citep{rillingBivariateEmpiricalMode2007}, T-EMD \citep{rehmanEmpiricalModeDecomposition2010}, and MEMD \citep{rehmanMultivariateEmpiricalMode2010}. MEMD is widely used in financial forecasting and forms Multivariate Decomposition Ensemble models (M-DEMs). \citet{dengMultistepaheadStockPrice2022} proposed the MEMD-LSTM model, combining MEMD for multidimensional time-frequency analysis and feature extraction with OATM-optimized LSTM hyperparameters. Experiments show that its multistep prediction surpasses EMD-LSTM and traditional LSTM. \citet{zouForecastingCrudeOil2022} developed a MEMD-CNN multiscale model for risk prediction, outperforming single-scale models like ARMA-GARCH and VMD-CNN. \citet{baiIntelligentForecastingModel2023} built a hybrid prediction model blending "fuzzy computing" and "decomposition-ensemble". Using MPE for feature selection, MI and NRS for filtering, and MEMD for decomposition, features entered an LSTM network. Though capturing market complexity, it struggles with unexpected events and has prediction delays. \citet{yaoStockIndexForecasting2023} proposed a hybrid MEMD-TCN model for stock index prediction, decomposing multidimensional data like opening, high, low, closing prices and trading volumes. The model outperforms univariate DEM and M-DEM combined with recurrent neural networks (RNNs) in stock index forecasting in several countries. It also emphasizes the superiority of TCNs compared to rnns.

However, MEMD suffers from computational inefficiency when dealing with multivariate data. \citet{langFastMultivariateEmpirical2018} proposed the FMEMD algorithm, which is far more computationally efficient than MEMD when dealing with large-scale multidimensional signals. \citet{thirumalaisamyFastAdaptiveEmpirical2018} proposed the FAMVEMD algorithm, which is much more practical when dealing with one-dimensional and three-dimensional synthetic signals. In addition to being computationally slow, \citet{caiMAEMDAlignedEmpirical2025}, standing for the prediction task, also pointed out that MEMD suffers from relaxing the quality constraints on subsequences, generating unnecessary frequency components, which negatively affects the final prediction performance. For this reason, they proposed Multivariate Aligned Empirical Mode Decomposition (MA-EMD). The algorithm first applies the standard EMD to each variable individually and then takes the IMFs of the target variable as a reference, uses the Kullback-Leibler dispersion (KLD) to measure the frequency similarity between the IMFs of different variables and aligns them to the corresponding sub-prediction models. It is experimentally demonstrated that M-DEM combined with MA-EMD outperforms the MEMD-based model regarding prediction accuracy.

In addition, reducing noise in financial data is another effective strategy to improve prediction results. However, existing DEM and M-DEM frameworks seldom consider the effects of noise when predicting financial data \citep{wangTwoStageDeepEnsemble2024}. Singular Spectrum Analysis (SSA) \citep{broomheadExtractingQualitativeDynamics1986} is a powerful noise reduction algorithm that is very effective in financial forecasting tasks. \citet{lahmiriMinuteaheadStockPrice2018} proposed an intraday stock price prediction method based on SSA and Support Vector Regression (SVR) combined with Particle Swarm Optimization (PSO).  Experimental results show that the method outperforms traditional prediction models in terms of accuracy. \citet{tangPredictionFinancialTime2021} used wavelet transform (WT) and SSA to reduce the raw data and then input the smoothed sequences into LSTM for prediction. This approach resulted in better prediction performance than feeding the original sequences directly into LSTM. \citep{wangDeepLearningIntegrated2024} proposed a hybrid deep learning framework SSA-PSO-LSTM for predicting stock index prices and their fluctuations. Experimental results show that this method has higher prediction accuracy during high volatility than the remaining models, such as PSO-LSTM. 

In summary, existing research has achieved some results in generating investor views and applying M-DEM, but there is still room for improvement. This paper proposes a hybrid SSA-MAEMD-TCN model which provides a new method for generating investor opinions and fills the research gap of M-DEMs in this field.

\section{Technical Background}
\subsection{Black-Litterman Model}
\begin{figure*}[htbp]
    \centering
    \includegraphics[width=0.9\textwidth]{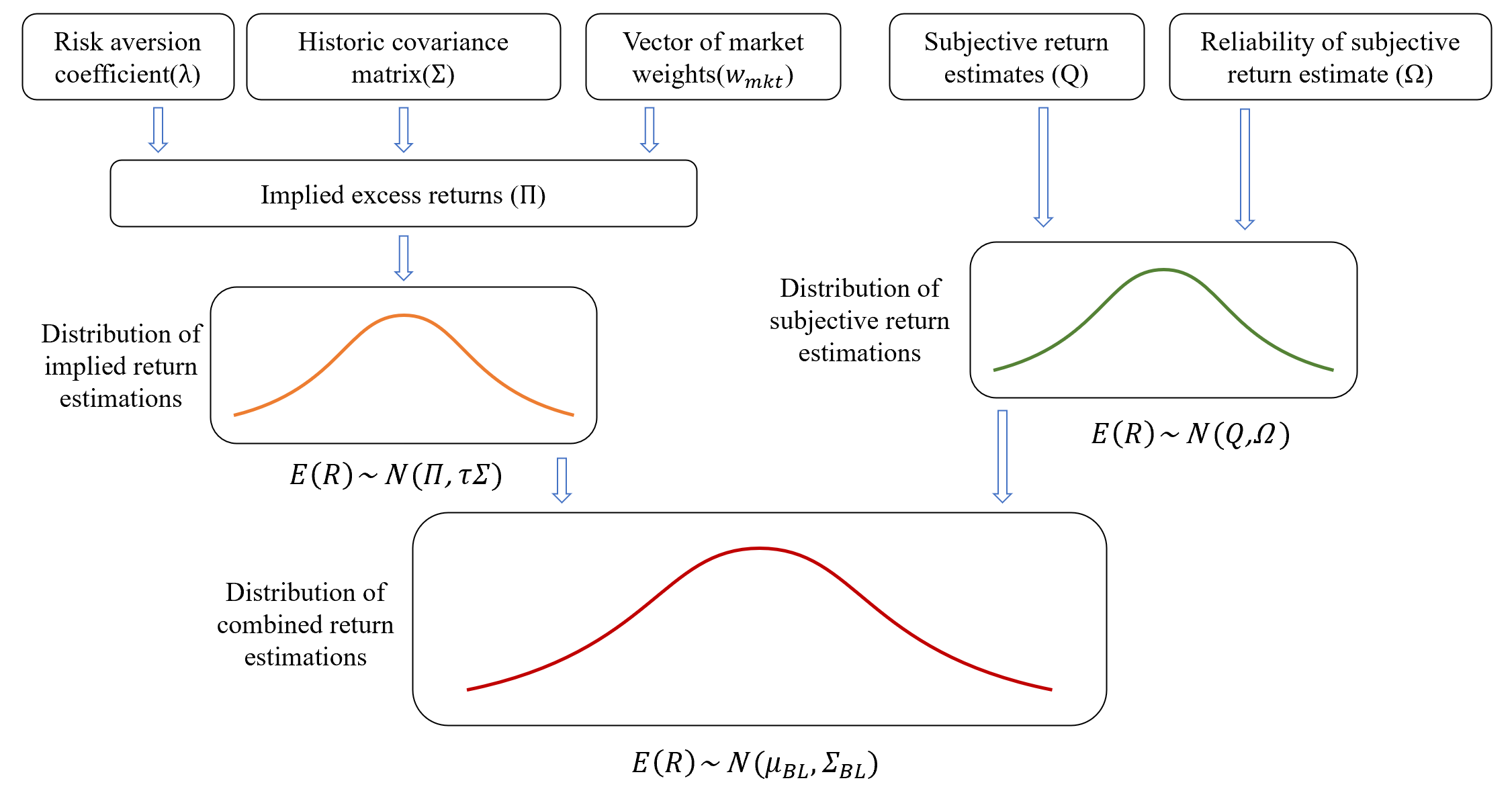}
    \caption{The process of Black-Litterman model \citep{idzorek2StepbystepGuide2007}.}
    \label{fig:bl}
\end{figure*}

To solve the problems of input sensitivity and estimation error maximization in traditional mean-variance optimization, Fischer Black and Robert Litterman developed the Black-Litterman model. Fig.~\ref{fig:bl} illustrates the key parameters and process of the model.

\subsubsection{Calculating Priori Returns:}
Starting from market equilibrium conditions, prior estimates of asset expected returns are derived through reverse optimization. Specifically, assuming that the market portfolio weights are based on market capitalization, the implied equilibrium returns are obtained using the reverse process of mean-variance optimization, as shown below:
\begin{equation}
\Pi = \lambda \Sigma w_{mkt}
\label{eq:fanxiang}
\end{equation}
where, $\Pi$ is the implied excess return vector ($N \times 1$), $\lambda$ is the risk aversion parameter. In this paper, we set \( \lambda = 2.5 \) to represent the world average risk aversion index. $\Sigma$ is the covariance matrix of excess returns ($N \times N$), $w_{mkt}$ is the vector of asset weights based on market capitalization ($N \times 1$).

\subsubsection{Setting Investor Views:}

Investor views are treated as new information, which can be either absolute or relative. These views are then converted into three parameters.

\textbf{View Matrix \( P \)}:  
The view matrix \( P \) is a \( k \times n \) matrix, where \( k \) is the number of views and \( n \) is the number of assets. It maps the investor’s views onto specific assets, indicating which assets correspond to each view.
Since we are only considering absolute views, each row of the \( P \) matrix will have exactly one non-zero element, which is located in the column corresponding to the asset and equals 1. Specifically, suppose you have \( k \) absolute views, corresponding to assets \( i_1, i_2, \ldots, i_k \) (where \( i_j \) is the index of the asset, and \( 1 \leq i_j \leq n \)), then the matrix \( P \) can be represented as:
\begin{equation}
        P = \begin{bmatrix}
            0 & 0 & \cdots & 1 & \cdots & 0 \\
            0 & 0 & \cdots & 1 & \cdots & 0 \\
            \vdots & \vdots & \ddots & \vdots & \ddots & \vdots \\
            0 & 0 & \cdots & 1 & \cdots & 0 \\
            \end{bmatrix}
        \label{eq:p}
        \end{equation}

\textbf{View Return Vector \( Q \)}:  
This is a \( k \times 1 \) vector representing the specific expected returns for each view. It captures the expected return for each individual view from the investor’s perspective. If we have forecasted the returns of \( k \) stocks to be \( R_1, R_2, \ldots, R_k \), the \( Q \) matrix can be represented as:
    \begin{equation}
        Q = \begin{bmatrix}
        R_1 \\
        R_2 \\
        \vdots \\
        R_k
        \end{bmatrix}
        \label{eq:q}
    \end{equation}
        
\textbf{View Error Covariance Matrix \( \Omega \)}:  
The view error covariance matrix \( \Omega \) is a \( k \times k \) diagonal matrix representing each view's uncertainty or confidence level. The elements along the diagonal represent the variance for each view, reflecting the investor's confidence in that view. When setting up the  \( \Omega \) matrix, we need to determine a scalar  \( \tau \). In this example, we use 500 days of historical data for calculating the assets' priori returns and priori covariances and therefore choose  \( \tau \) = 1/500.
\begin{equation}
\Omega = \begin{bmatrix}
    (p_1 \Sigma p_1') \tau & 0 & \cdots & 0 \\
    0 & (p_2 \Sigma p_2') \tau & \cdots & 0 \\
    \vdots & \vdots & \ddots & \vdots \\
    0 & 0 & \cdots & (p_k \Sigma p_k') \tau
    \end{bmatrix}
    \label{eq:omega}
    \end{equation}
where, \( p_i \) is the asset weight vector for the i-th view, reflecting the investor's assessment of that asset's importance or weight in the view. \( \Sigma \) is the covariance matrix of asset returns, indicating the correlations and volatilities of the asset returns.

\subsubsection{Calculating Posterior Returns:}  

The priori return distribution and the subjective view distribution are combined using bayesian methods to calculate the posterior estimate of asset expected returns, as shown below:
\begin{equation}
\label{eq:u_post}
\mu_{post} = \left( (\tau \Sigma)^{-1} + P^T \Omega^{-1} P \right)^{-1} \left( (\tau \Sigma)^{-1} \Pi + P^T \Omega^{-1} Q \right)
\end{equation}

And the formula for the posterior covariance ($\Sigma_{post}$) is:
\begin{equation}
\label{eq:xfc_post}
\Sigma_{post} = \Sigma + \left( (\tau \Sigma)^{-1} + P^T \Omega^{-1} P \right)^{-1}
\end{equation}

\subsubsection{Optimizing Asset Allocation:}  

The posterior returns and the posterior covariance matrix are input into the mean-variance model for optimization to obtain the specific asset allocation weights \( w_{bl} \), as shown below:
\begin{equation}
\label{eq:w_bl}
w_{bl} = \left( \lambda \Sigma_{post} \right)^{-1} \mu_{post}
\end{equation}

\subsection{Multivariate Aligned Empirical Mode Decomposition (MA-EMD)} 
MA-EMD is an improved version of MEMD, aiming to solve the problems in multivariate decomposition, such as degradation of decomposition quality and computational inefficiency. In MA-EMD, each variable is independently decomposed by the standard EMD to produce a series of IMFs. However, EMD leads to different IMFs for eachvariable, which the KLD-based frequency alignment module can solve. The main task of this module is to use the extreme value intervals of IMFs to characterize their frequency properties and then measure the similarity of the two IMFs using the Kullback-Leibler Divergence (KLD). Finally, frequency alignment is achieved by assigning each IMF of the related variable (OHLV) to the IMF of the target variable (Close) with the smallest KLD.

\subsubsection{EMD Algorithm}
Empirical Mode Decomposition (EMD) is an adaptive signal processing technique suitable for nonlinear and non-stationary signals. Its core is decomposing complex signals into a set of IMFs with different time scales. The first step is to determine the local extreme values of the signal x(t) and use spline interpolation to generate the upper envelope u(t) and the lower envelope l(t). Then calculate the average value m(t) of the two envelopes:
\begin{equation}
m(t) = \frac{u(t) + l(t)}{2}
\end{equation}
Following this, the mean envelope is subtracted from the original signal to obtain the intrinsic mode function (IMF):
\begin{equation}
h(t) = x(t) - m(t)
\end{equation}

A qualified IMF must verify that h(t) satisfies two conditions. First, the meaning of the upper and lower envelopes must be zero. Second, the difference between the number of extreme values and the number of zero crossings of h(t) must not exceed 1. If these two conditions are not met, the iterative decomposition process continues until the conditions are met. After the decomposition is completed, the original signal x(t) can be expressed as the sum of multiple IMFs and a residual term.
\begin{equation}
x(t) = \sum_{i=1}^{n} h_i(t) + r_n(t)
\end{equation}

\subsubsection{Frequency Alignment Module}
The algorithm uses the spacing between extreme values as a frequency representation method. The extreme value spacing distribution is sufficient for high-frequency IMFs. However, as the extreme values become more and more sparse, the applicability of this indicator for low-frequency IMFs gradually decreases. \citet{caiMEDEMMNNbasedEmpirical2024} concluded that partial decomposition does not degrade the performance of the decomposition ensemble model. Therefore, the MA-EMD algorithm stops the decomposition process by setting a threshold \( \omega \) of the number of extreme values. In this paper, we take \( \omega = 20 \).

The module uses Kullback-Leibler divergence (KLD) to measure the similarity between the distributions of the extreme value intervals of two IMFs. Before calculating the KLD value, the probability distribution of each IMF needs to be expanded to include all possible extreme value intervals for each IMF. Then, Laplace smoothing is applied to ensure that the probabilities are not zero. The calculation formula is as follows:
\begin{equation}
P'(X = x) = 
\begin{cases} 
  \frac{N_x}{N - 1} + \epsilon, & \text{if } x \in \mathcal{X} \\
  \epsilon, & \text{if } x \in \mathcal{X}' \setminus \mathcal{X}
\end{cases}  
\label{eq:extend the probability distribution}
\end{equation}
where: \( N_x \) refers to the number of occurrences of the mechanism interval \( x \) in the IMF; \( N \) is the total number of extrema points; \( \mathcal{X} \) is the set of extrema intervals that exist in this particular IMF; \( \mathcal{X}' \) represents the set of extended extrema intervals across all IMFs; \(\epsilon\) is a very small constant, \(\epsilon\) = 0.000001. 

The final probability distribution \(P''(X = x)\) and the formula for calculating the similarity between distributions using KLD are given by Eqs. \ref{eq:prob_distribution} and \ref{eq:kld}, respectively.
\begin{equation}
\label{eq:prob_distribution}   
P''(X = x) = \frac{P'(X = x)}{\sum_{x \in \mathcal{X}'} P'(X = x)}
\end{equation}
\begin{equation}
\label{eq:kld}
D_{KL}(P'' \parallel Q'') = \sum_{x \in \mathcal{X}} P''(x) \log \frac{P''(x)}{Q''(x)}
\end{equation}
where \( P\) and \(Q\) represent the probability distributions for the related variable and the target variable, respectively, and \( \mathcal{X} \) is the extended range of the two distributions.

\subsection{Temporal Convolutional Network (TCN)}
TCN is a deep learning model specialized in processing time series data. Its design is inspired by Convolutional Neural Networks (CNNs) structure and optimized and improved for time series prediction tasks. The core innovation of TCN is the combination of dilated convolution and causal convolution techniques.

\subsubsection{Causal Convolution}
Causal convolution is one of the core components of TCN, ensuring that the network uses only the current and past information for predictions, without leaking any future information. In causal convolution, the convolution operation is performed only on the current time step and the data before it, achieved by adding zero-padding to the input data. For a one-dimensional sequence input $x \in \mathbb{R}^n$ and a convolutional filter $f : \{0, \ldots, k-1\} \to \mathbb{R}$, the causal convolution operation $F$ at the $s$-th element of the sequence is defined as:
\begin{equation}
      F(s) = (x * f)(s) = \sum_{i=0}^{k-1} f(i) \cdot x_{s-i}  
\end{equation}
where, $k$ is the size of the convolutional kernel, and $s$ is the index of the current time step. This convolution ensures that the output $F(s)$ depends only on $x_0, \ldots, x_s$, and not on future inputs $x_{s+1}, \ldots, x_n$.

\begin{figure}[htbp]

    \centering
    \includegraphics[width=0.45\textwidth]{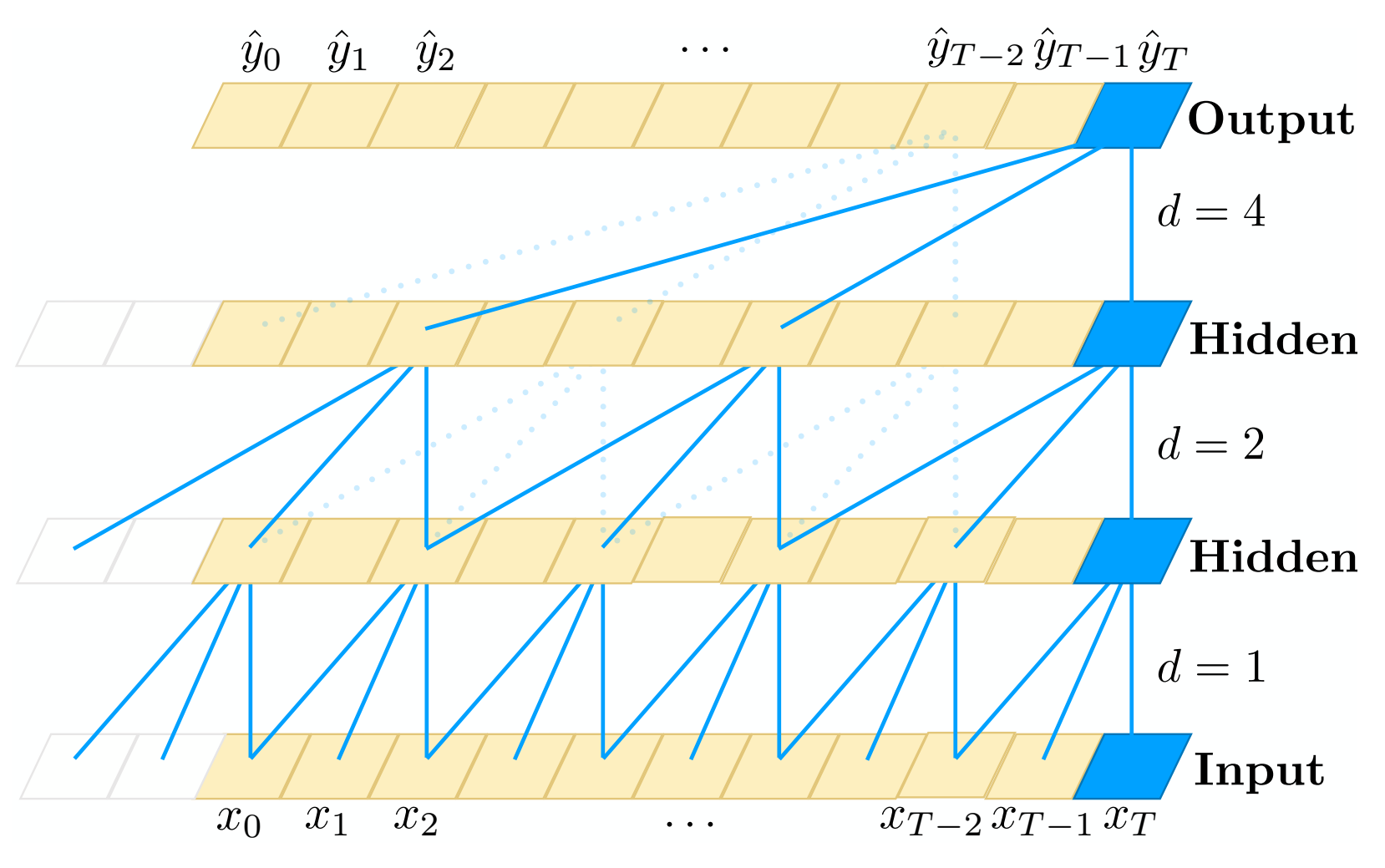}
    \caption{A dilated causal convolution with dilation factors d = 1, 2, 4 and filter size k = 3 \citep{baiEmpiricalEvaluationGeneric2018}.}
    \label{fig:fig1}

\end{figure}

\subsubsection{Dilated Convolution}

Dilated convolution introduces a skip interval in the convolutional kernel, allowing the receptive field of TCN to grow exponentially. For example, when the dilation factor is 2, the convolution kernel skips one data point during the convolution operation. As the network depth increases, the dilation factor gradually increases, enabling the network to capture longer temporal dependencies, as shown in Fig.~\ref{fig:fig1}. For a dilation factor $d$, the dilated convolution operation $F_d$ at the $s$-th element of the sequence is defined as:
\begin{equation}
F_d(s) = (x *^d f)(s) = \sum_{i=0}^{k-1} f(i) \cdot x_{s-d \cdot i}  
\end{equation}
where $d$ is the dilation factor, and $k$ is the size of the convolutional kernel.

\subsubsection{Residual Connections}
\begin{figure*}[htbp]
    \centering
    {\includegraphics[width=0.4\textwidth]{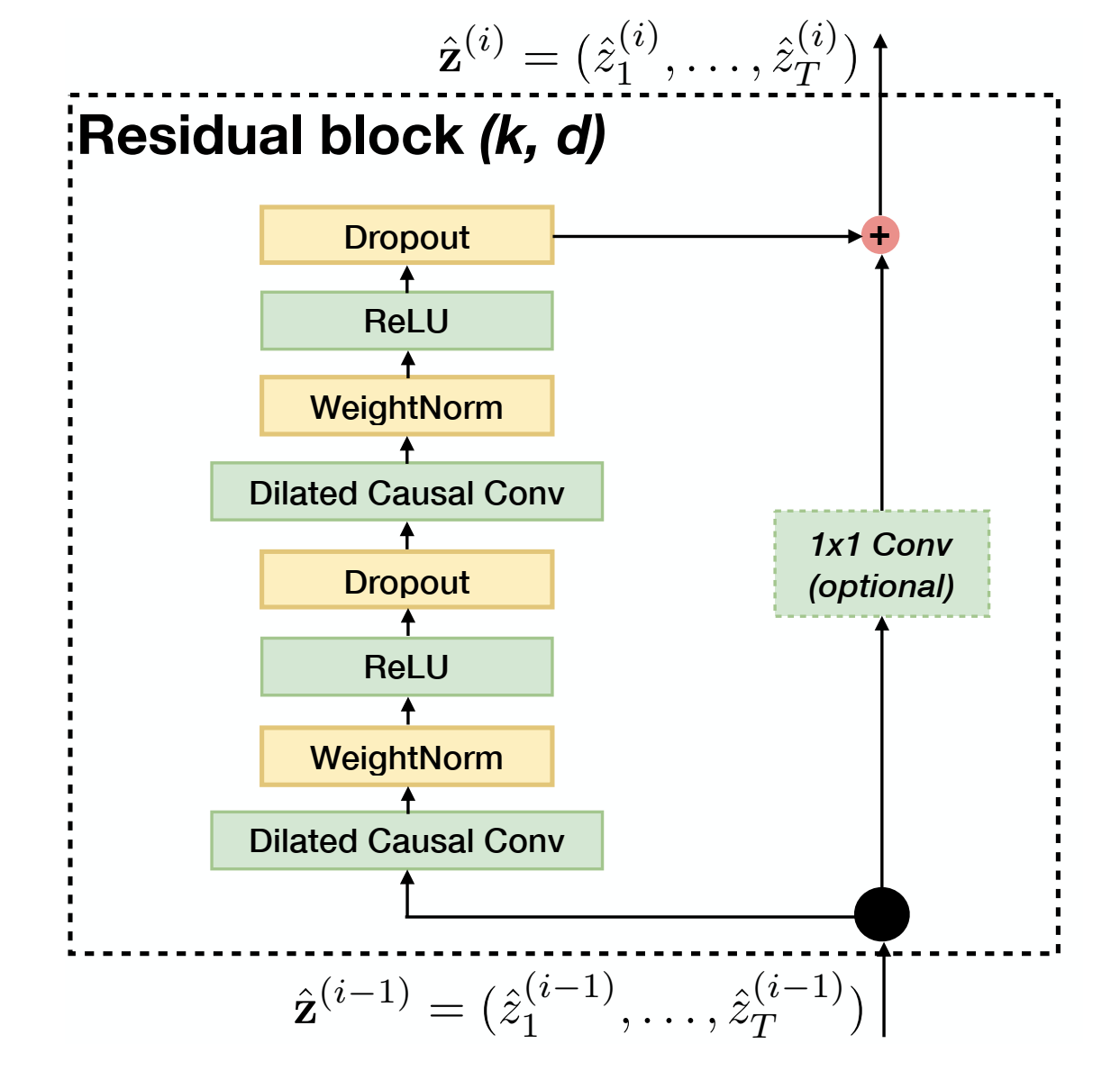}\label{fig:left}}
    \hspace{0.1cm} 
    {\raisebox{0.5cm}{\includegraphics[width=0.4\textwidth]{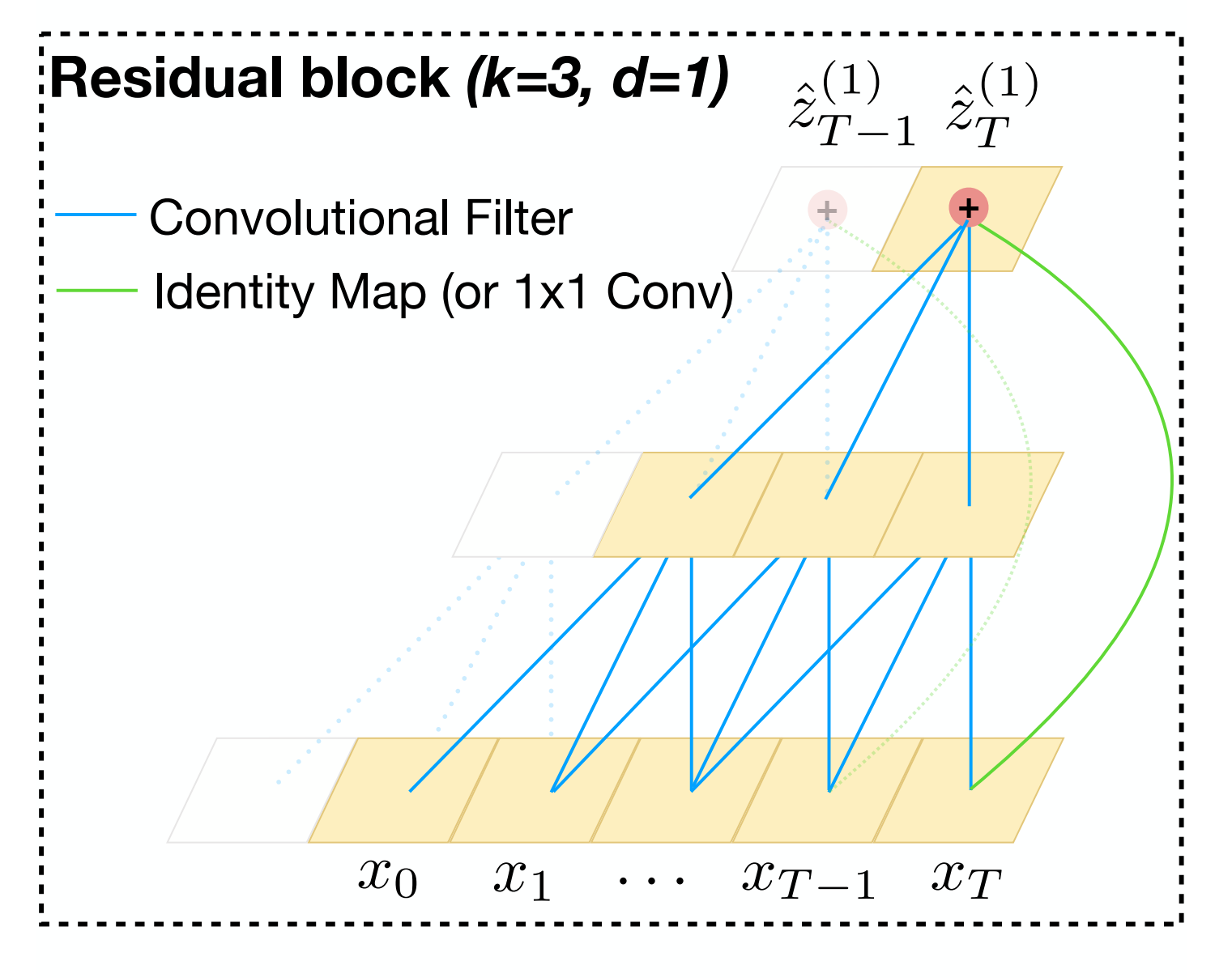}\label{fig:right}}}
    \caption{The residual block in TCNs. Left plot: an \( 1 \times 1 \) convolution is added when the residual input and output have different dimensions. Right plot: an example of the residual connection in TCN \citep{baiEmpiricalEvaluationGeneric2018}. }
    \label{fig:merged}
\end{figure*}

To alleviate the issues of vanishing and exploding gradients during the training of deep networks, TCN introduces residual connections. Residual connections make it easier for the network to learn identity mappings by directly adding the input to the output. In TCN, each residual block typically consists of two dilated convolution layers, weight normalization, ReLU activation functions, and a dropout layer. If the dimensions of the input and output are not consistent, a 1x1 convolution can be used to adjust the dimensions to allow for element-wise addition. As shown in Fig.~\ref{fig:merged}.
For an input $x$ and the output of a convolutional layer $F(x)$, the output of a residual block $o$ is:
\begin{equation}
    o = \text{Activation}(x + F(x))
\end{equation}
where $\text{Activation}$ is a nonlinear activation function, such as ReLU. 

\subsection{Singular Spectral Analysis (SSA)}
SSA is a classic method for processing time series data. It embeds data into a high-dimensional space and uses a sliding window combined with singular value decomposition (SVD) to extract the main patterns.

The first step of SSA involves reconstructing the time series into a matrix containing lagged vectors. Starting with a time series of length \( N \), denoted as \( \{ x_1, x_2, ..., x_N \} \), a window size \( L \) is selected, typically much smaller than \( N \) (usually \( L < N/2 \)). This window is used to construct a trajectory matrix \( X \), which has \( L \) rows and \( K = N - L + 1 \) columns. The matrix \( X \) is built by sliding the window across the time series, with each row representing a "snapshot" of the series at a given time point. The structure of the time series over time is thus captured in this matrix:
\begin{equation}
 X = 
\begin{bmatrix}
x_1 & x_2 & \dots & x_K \\
x_2 & x_3 & \dots & x_{K+1} \\
\vdots & \vdots & \ddots & \vdots \\
x_{L} & x_{L+1} & \dots & x_N
\end{bmatrix}   
\end{equation}

With the trajectory matrix in hand, the next step is to decompose it further using SVD. SVD is a technique that breaks down the matrix into three parts: two orthogonal matrices and a diagonal matrix containing the singular values. The SVD of matrix \( X \) is expressed as:
\begin{equation}
X = U \Sigma V^T 
\end{equation}
where \( U \in \mathbb{R}^{L \times L} \) contains the left singular vectors, \( \Sigma \in \mathbb{R}^{L \times K} \) is the singular value matrix with diagonal values representing the strength of each component, and \( V^T \in \mathbb{R}^{K \times K} \) contains the right singular vectors.

Following the decomposition, the time series is divided into multiple linearly independent subseries through grouping. The singular values \( \{\sigma_1, \sigma_2, \ldots, \sigma_r\} \) are partitioned into \( m \) groups \( \{I_1, I_2, \ldots, I_m\} \), with each group corresponding to a submatrix. For the \( i \)-th group, the submatrix \( X_i \) is formed by multiplying the corresponding singular values, left singular vectors, and right singular vectors:
\begin{equation}
    X_i = \sum_{j=1}^{k_i} \sigma_{i_j} u_{i_j} v_{i_j}^T
\end{equation}
where \( u_{i_j} \) and \( v_{i_j} \) are the \( i_j \)-th columns of \( U \) and \( V \), respectively.

The final step involves reconstructing the time series from the grouped submatrices. Anti-diagonal averaging is performed on each submatrix to obtain the reconstructed series. For each group \( \{I_1, I_2, \ldots, I_m\} \), the corresponding submatrix \( X_i \) is averaged along the anti-diagonals to produce a sequence \( \hat{y}_i \) of length \( N = L + K - 1 \). The formula for anti-diagonal averaging is:
\begin{equation}
     \hat{y}_i(k) = \frac{1}{n_k} \sum_{j=1}^{n_k} X_i(j, k-j+1)
\end{equation}
where \( n_k \) is the number of elements on the \( k \)-th anti-diagonal. The reconstructed original time series \( \hat{Y}_N \) is obtained by summing these sequences:
\begin{equation}
    \hat{Y}_N = \sum_{i=1}^{m} \hat{y}_i
\end{equation}
This reconstruction process combines the decomposed subseries to produce a denoised version of the original time series.

\begin{table*}[ht]
\centering
\begin{minipage}[t]{0.45\textwidth}
\caption{Stock Pool for Portfolio Construction}
\label{tab:stock_pool}
\renewcommand{\arraystretch}{1.2} 
\setlength{\tabcolsep}{2pt} 

\resizebox{1.05\textwidth}{!}{
\begin{tabular}{cccc}
\toprule
\textbf{Stock} & \textbf{Sector} & \textbf{Stock} & \textbf{Sector} \\
\midrule
AAPL & Technology & COST & Consumer Discretionary \\
MSFT & Technology & AZN & Health Care \\
GOOG & Technology & AMGN & Health Care \\
GOOGL & Technology & ISRG & Health Care \\
NVDA & Technology & PEP & Consumer Staples \\
INTC & Technology & HON & Industrials \\
AMD & Technology & LIN & Industrials \\
AMZN & Consumer Discretionary & TMUS & Telecommunications \\
TSLA & Consumer Discretionary & CMCSA & Telecommunications \\
NFLX & Consumer Discretionary & CSCO & Telecommunications \\
\bottomrule
\end{tabular}}
\end{minipage}
\hspace{0.08\textwidth} 
\begin{minipage}[t]{0.45\textwidth}
\centering
\renewcommand{\arraystretch}{1.32} 
\setlength{\tabcolsep}{3pt} 
\caption{The statistical analysis results of selected stocks}
\label{tab:table_2}
\footnotesize
\begin{tabular}{ccccc}
\toprule
\textbf{Stock} & \textbf{Max} & \textbf{Min} & \textbf{Mean} & \textbf{Std} \\
\midrule
AAPL & 194.7577 & 54.4499 & 134.6913 & 32.1692 \\
MSFT & 354.6355 & 129.6211 & 248.1334 & 50.0254 \\
GOOGL & 149.1255 & 52.4557 & 104.9878 & 24.7227 \\
NVDA  & 49.3326 & 4.8924 & 18.8433 & 9.4758 \\
TSLA   & 409.9700 & 24.0813 & 206.1424 & 88.8008 \\
AMGN   & 270.2472 & 156.2849 & 213.2018 & 18.7573 \\
PEP  & 185.9797 & 90.6220 & 144.5176 & 21.3784 \\
HON  & 217.2645 & 93.6586 & 179.9061 & 25.2358 \\
\bottomrule
\end{tabular}
\end{minipage}
\end{table*}

\section{Experiment design}

\subsection{Data description}
To construct an investment portfolio, we carefully selected 20 representative stocks from various industries within the NASDAQ 100 index as the targets for our portfolio. These stocks cover key areas such as technology, healthcare, and consumer sectors and are known for their high growth potential and innovative capabilities, as shown in Table~\ref{tab:stock_pool}. We collected the daily trading data of these stocks from January 1, 2020, to August 31, 2023, from Yahoo Finance. 

Among the 20 stocks, we focus on eight key stocks, including Apple (AAPL), Microsoft (MSFT), Alphabet Class A (GOOGL), Nvidia (NVDA), Tesla (TSLA), Amgen (AMGN), PepsiCo (PEP), and Honeywell International (HON). These stocks hold significant market positions in their respective sectors, and detailed statistical analyses are presented in Table~\ref{tab:table_2}.

\subsection{Proposed Methodology}
\begin{figure*}[htbp]
    \centering
    \includegraphics[width=1\textwidth]{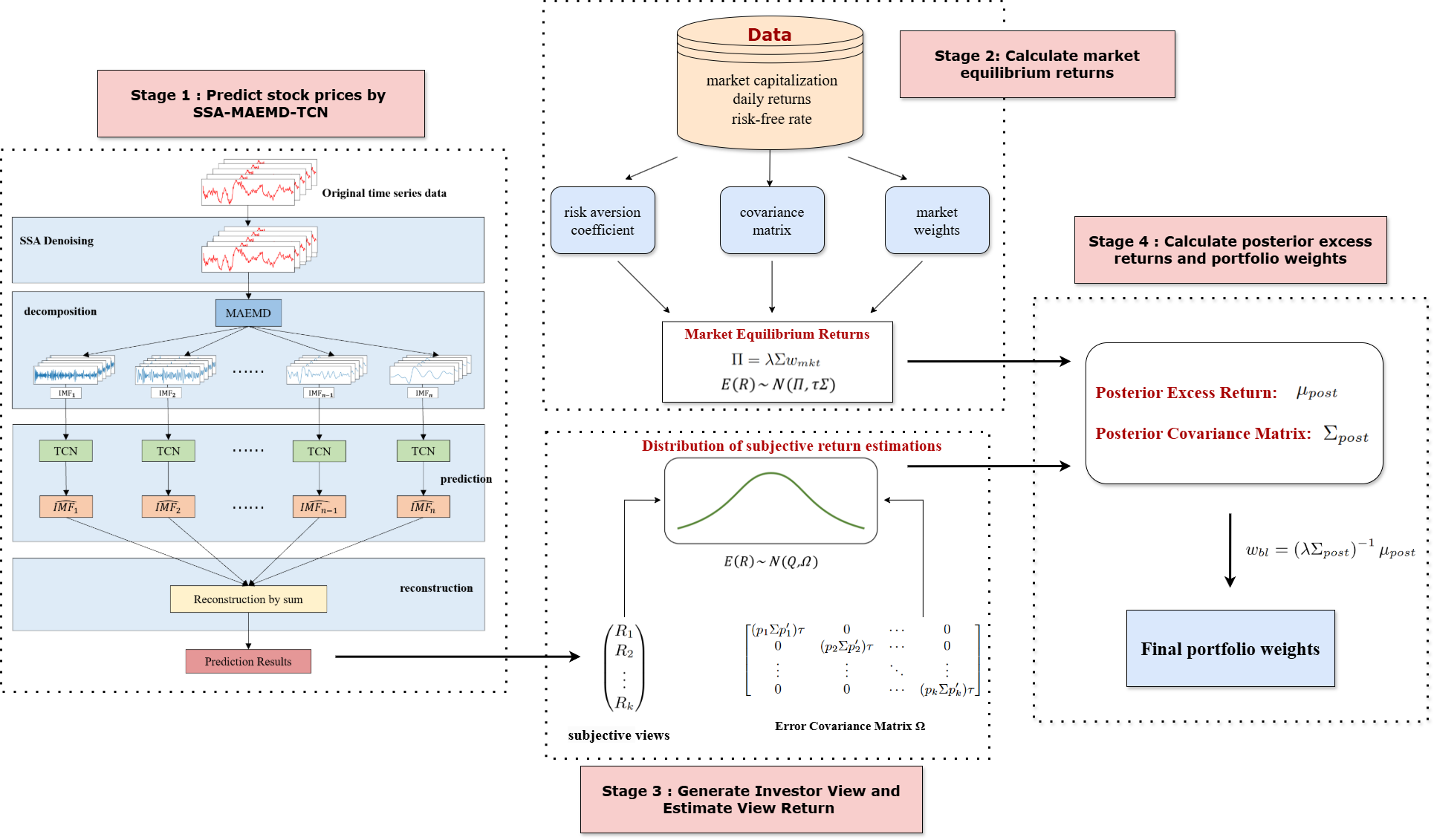}
    \caption{Procedure of the complete Black-Litterman strategy.}
    \label{fig:whole}
\end{figure*}

In this paper, we propose a novel stock price prediction model, SSA-MAEMD-TCN, and consider the model as a way to enhance the subjective view of the investor and thus optimize the performance of Black-Litterman's portfolio. See Fig. \ref{fig:whole} for the overall framework.

\noindent \textbf{(1) Predict stock prices:} 

\begin{figure*}[!ht]
    \centering

    \begin{minipage}{0.18\textwidth}
        \centering
        \text{\small Close}
        \includegraphics[width=\textwidth, height=0.3\textheight]{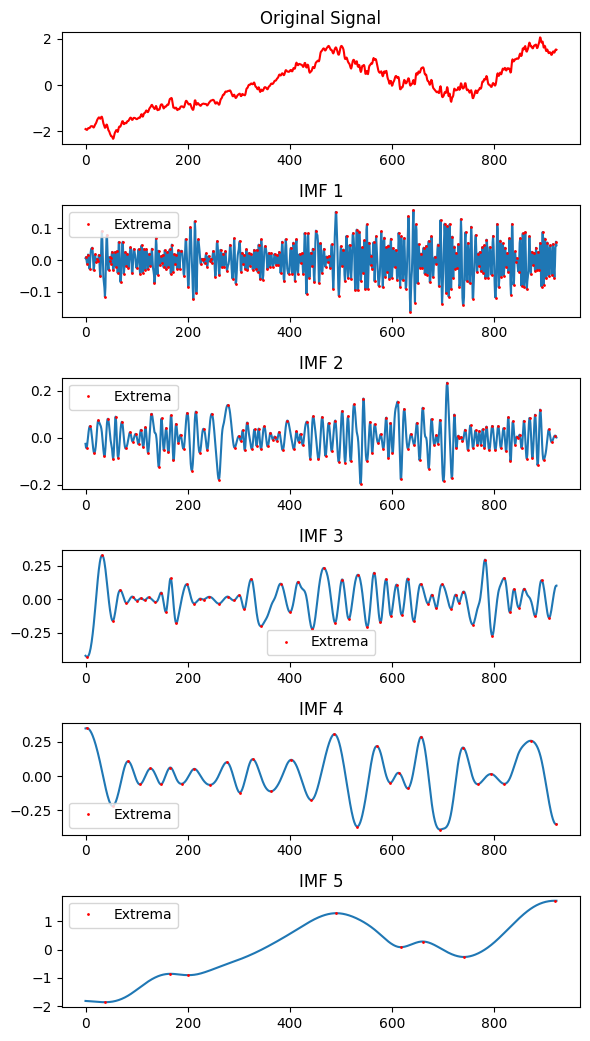}
    \end{minipage} \hspace{0.01\textwidth} 
    \begin{minipage}{0.18\textwidth}
        \centering
        \text{\small Open}
        \includegraphics[width=\textwidth, height=0.3\textheight]{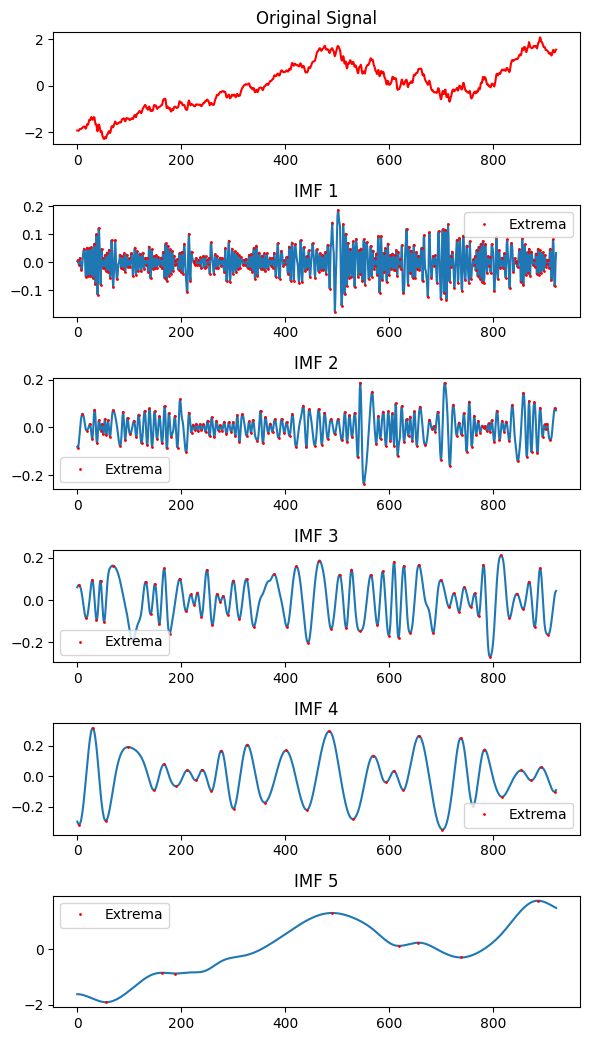}
    \end{minipage} \hspace{0.01\textwidth}
    \begin{minipage}{0.18\textwidth}
        \centering
        \text{\small High}
        \includegraphics[width=\textwidth, height=0.3\textheight]{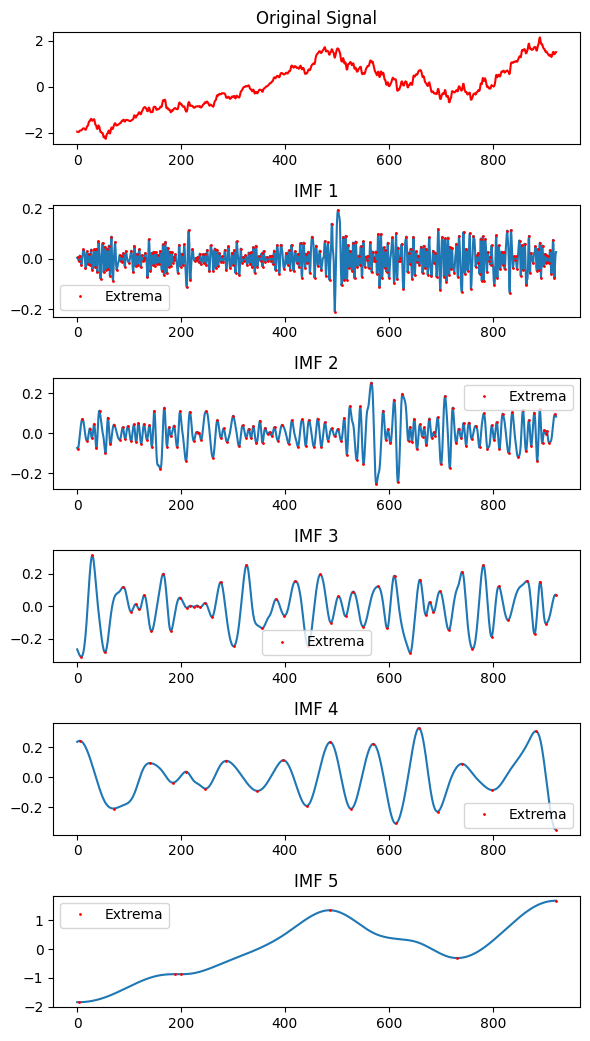}
    \end{minipage} \hspace{0.01\textwidth}
    \begin{minipage}{0.18\textwidth}
        \centering
        \text{\small Low}
        \includegraphics[width=\textwidth, height=0.3\textheight]{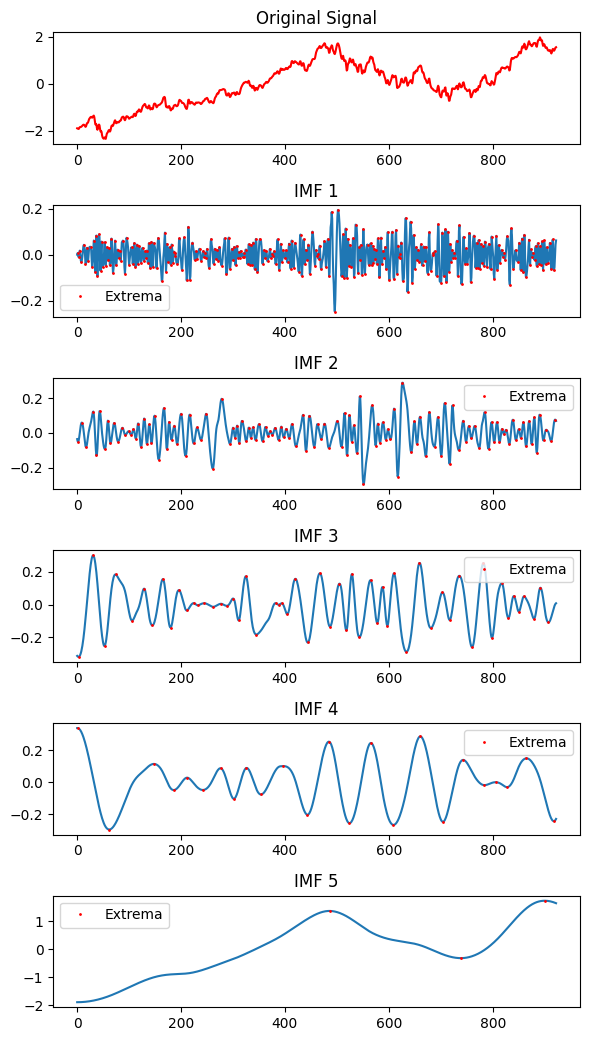}
    \end{minipage} \hspace{0.01\textwidth}
    \begin{minipage}{0.18\textwidth}
        \centering
        \text{\small Volume}
        \includegraphics[width=\textwidth, height=0.3\textheight]{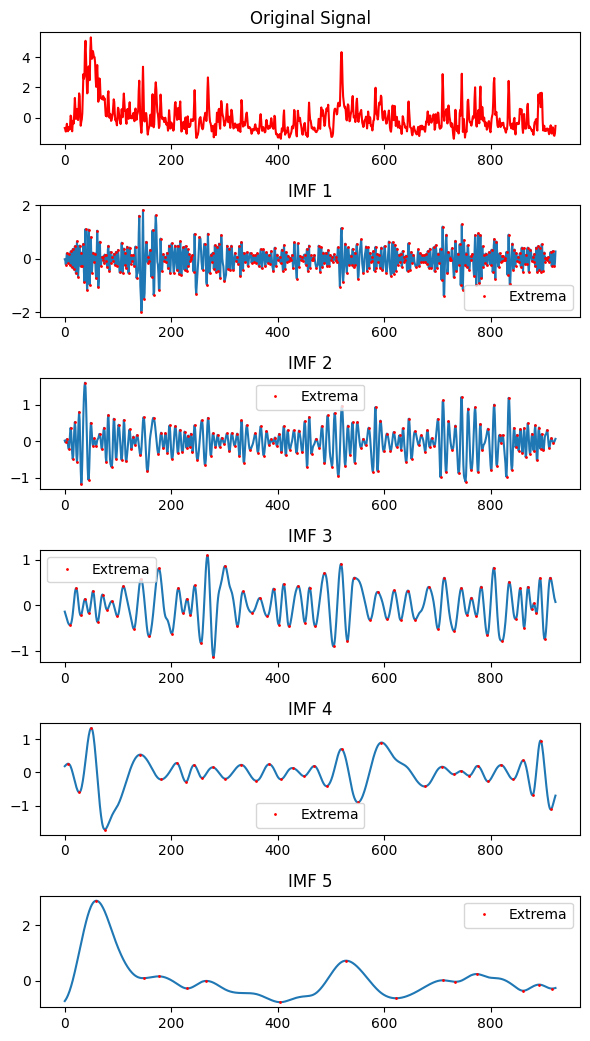}
    \end{minipage}

    \caption{COHLV of MSFT was decomposed using MA-EMD.}
    \label{fig:MAEMD}
\end{figure*}

This stage aims to predict stock prices using a hybrid model. First, the original time series data are subjected to SSA denoising to reduce the noise interference. The denoised data are then decomposed into multiple IMFs using MA-EMD, as shown in Fig.~\ref{fig:MAEMD}, where it can be visualized that the number of IMFs for each variable is equivalent. Before decomposition, normalizing the data is not negligible, as shown in the following equations:
\begin{equation}
z_{t, i} = \frac{x_{t, i} - \mu_i}{\sigma_i}
\end{equation}
Where, \( x_{t, i} \) is the raw value of the \( i \)-th feature at time \( t \). \( \mu_i \) is the mean of the \( i \)-th feature, \( \sigma_i \) is the standard deviation of the \( i \)-th feature.

The final step of the prediction framework is to predict the IMF for each alignment using a TCN. We use the sliding window method to divide the data into many short time intervals using a fixed-length seven-step input, with the eighth data point serving as a label. The window gradually moved forward until the last window reached. Regarding the division of the dataset, 70\% of the data is assigned to the training set, 15\% to the validation set, and the remaining 15\% as the test set (out-of-sample data). Finally, the predicted values of each IMF are summed up to give the final prediction. See Table \ref{tab:hyperparameters} for specific parameters of the model.

\begin{table}[h]
\centering

\setlength{\tabcolsep}{8pt}        
\caption{Hyperparameter tuning for TCNs.}
\label{tab:hyperparameters}
\begin{tabular}{cc}
\toprule
\textbf{Hyperparameter} & \textbf{Chosen Value} \\
\midrule
Kernel\_sizes & 2 \\
Hidden\_size & [64, 128] \\
Learning\_rate  & 0.001 \\
Layer\_num  & 2 \\
Dropout rate & [0, 0.3] \\
Epochs & [50, 100] \\
Batch\_size & [16, 32] \\
Activation function & ReLU \\
Optimizer & Adam \\
\bottomrule
\end{tabular}
\end{table}

\noindent \textbf{(2) Construct the portfolio:} 

The next step is to calculate the market equilibrium return based on inputs such as market capitalization, daily return rate, and risk-free rate.  Inverse optimization is used to obtain the equilibrium return after determining the risk aversion coefficient and constructing the covariance matrix. The asset prices predicted in the first stage are converted into returns to generate subjective opinions. Finally, the posterior excess return and posterior covariance are obtained according to Eqs.~\ref{eq:u_post} and \ref{eq:xfc_post}, and the optimized portfolio weights are obtained using Eq.~\ref{eq:w_bl}.

\subsection{Assessment metric}
\subsubsection{Predictive model evaluation metrics}

To evaluate the performance of the proposed hybrid model, we employ three metrics for assessment: RMSE, MAPE, and the R$^2$, based on Eqs. (\ref{eq:rmse}, \ref{eq:mape}, \ref{eq:r2}). RMSE measures the difference between the predicted and actual values; the smaller the value, the closer the model's prediction is to the exact value. MAPE expresses the error as a percentage, mainly reflecting the difference between the predicted and actual values. R² measures the model's goodness of fit, and its value ranges from 0 to 1. The closer the value of R² to 1, the better the model's fit to the data.

\begin{equation}
\label{eq:rmse}
\text{RMSE} = \sqrt{\frac{1}{N} \sum_{t=1}^{N} (y_t - \hat{y}_t)^2}
\end{equation}

\begin{equation}
\label{eq:mape}
\text{MAPE} = \frac{1}{N} \sum_{t=1}^{N} \left|\frac{y_t - \hat{y}_t}{y_t}\right| \times 100\%
\end{equation}

\begin{equation}
\label{eq:r2}
R^2 = 1 - \frac{\sum_{t=1}^{N} (y_t - \hat{y}_t)^2}{\sum_{t=1}^{N} (y_t - \bar{y})^2}
\end{equation}

In the aforementioned formula, \( y_t \) denotes the actual value, \( \hat{y}_t \) denotes the predicted value,  \( \bar{y} \) is the mean of the actual values, and \( N \) represents the total number of samples.

\subsubsection{Portfolio evaluation metrics}

The Sharpe Ratio is a commonly used metric in finance for measuring the performance of investment portfolios. It assesses the risk-adjusted return of a portfolio, that is, the excess return per unit of total risk. 
\begin{equation}
\label{sharp}
\text{Sharpe Ratio} = \frac{R_p - R_f}{\sigma_p}
\end{equation}
where, \( R_p \) is the expected return of the portfolio. \( R_f \) is the risk-free rate. \( \sigma_p \) is the standard deviation of the portfolio's returns, representing the total risk.

The Herfindahl-Hirschman Index (HHI) is used to quantify the concentration and diversification of a portfolio. The HHI reflects the concentration of a portfolio by adding the squares of the weights of each asset in the portfolio. The lower the HHI value, the more diversified the portfolio is.
\begin{equation}
\text{HHI} = \sum_{i=1}^{n} (w_i)^2
\end{equation}
where \( w_i \) represents the weight of the \( i \)-th stock, and \( n \) is the total number of stocks in the market.

\section{Result Discussion}
\subsection{Prediction results}
In this section, we perform a comprehensive comparative analysis of the proposed SSA-MAEMD-TCN hybrid model with the benchmark models (MAEMD-TCN, MEMD-TCN, MAEMD-LSTM) to evaluate their performance in stock price prediction tasks. We used three evaluation metrics: RMSE, MAPE and R$^2$. These metrics can reflect the predictive accuracy and goodness-of-fit of the models from different perspectives.

\begin{table*}[htbp]

    \centering
     
    \begin{minipage}{0.45\textwidth}
    \renewcommand{\arraystretch}{1.1}

    \setlength{\tabcolsep}{1.5pt} 
        \caption{Prediction accuracy of different models.}
        \label{tab:results}
        \footnotesize
        \begin{tabular}{lccccc}
            \toprule
    Stock & metric & {Model-1\footnotemark[1]} & {Model-2\footnotemark[2]} & {Model-3\footnotemark[3]} & {Model-4\footnotemark[4]}\\
    \midrule
    AAPL  & RMSE & 1.5629 & 1.9308 & 2.3602 & 2.9410\\
          & MAPE & 0.0076 & 0.0089 & 0.0109 & 0.0125\\
          & R$^2$  & 0.9866 & 0.9796 & 0.9674 & 0.9491\\
          
    MSFT  & RMSE & 2.9177 & 3.3560 & 4.4554 & 5.4004\\
          & MAPE & 0.0078 & 0.0083 & 0.0122 & 0.0139\\
          & R$^2$& 0.9899 & 0.9866 & 0.9765 & 0.9652\\
          
    GOOGL  & RMSE & 1.1355 & 1.7628 & 2.1211 & 1.9092\\
          & MAPE & 0.0080 & 0.0119 & 0.0154 & 0.0137\\
          & R$^2$& 0.9919 & 0.9805 & 0.9716 & 0.9772\\
          
    NVDA  & RMSE & 0.8591 & 0.9658 & 1.3608 & 1.5789\\
          & MAPE & 0.0215 & 0.0218 & 0.0252 & 0.0322\\
          & R$^2$& 0.9899 & 0.9876 & 0.9753 & 0.9624\\
          
    TSLA  & RMSE & 4.4226 & 5.6694 & 5.4443 & 5.9504\\
          & MAPE & 0.0138 & 0.0206 & 0.0220 & 0.0221\\
          & R$^2$& 0.9874 & 0.9793 & 0.9809 & 0.9771\\
          
    AMGN  & RMSE & 1.4308 & 2.5348 & 2.9539 & 3.6569\\
          & MAPE & 0.0050 & 0.0088 & 0.0120 & 0.0129\\
          & R$^2$& 0.9866 & 0.9579 & 0.9429 & 0.9124\\
          
    PEP   & RMSE & 1.0452 & 2.1527 & 1.6807 & 2.2738\\
          & MAPE & 0.0047 & 0.0104 & 0.0080 & 0.0105\\
          & R$^2$& 0.9652 & 0.8533 & 0.9106 & 0.8226\\
          
    HON  & RMSE & 1.2415 & 1.3809 & 1.4816 & 1.8478\\
         & MAPE & 0.0050 & 0.0058 & 0.0063 & 0.0075\\
         & R$^2$& 0.9604 & 0.9508 & 0.9435 & 0.9122\\
\bottomrule
        \end{tabular}
        
        \vspace{0.3cm}
        \footnotesize{1:SSA-MAEMD-TCN, 2:MAEMD-TCN, 3:MEMD-TCN, 4:MAEMD-LSTM}
    \end{minipage}
    \hspace{0.05\textwidth} 
    \begin{minipage}{0.4\textwidth}
    \renewcommand{\arraystretch}{1.1}
    \setlength{\tabcolsep}{3pt}
        \caption{Learning ability of different models for high-frequency imfs.}\label{tab:performance_comparison}
          \footnotesize
          \renewcommand{\thefootnote}{\arabic{footnote}}
        \begin{tabular}{llcccccc}
            \toprule
            \multirow{2}{*}{Stock} & \multicolumn{3}{c}{SSA-MAEMD-TCN} & \multicolumn{3}{c}{MAEMD-TCN} \\
            \cmidrule(lr){2-4} \cmidrule(lr){5-7}
            & IMF & RMSE & R$^2$ & IMF & RMSE & R$^2$ \\
            \midrule
            \multirow{3}{*}{AAPL} 
              & IMF\_1 & 0.0155 & 0.8530 & IMF\_1 & 0.0369 & -0.4201 \\
              & IMF\_2 & 0.0052 & 0.9921 & IMF\_2 & 0.0122 & 0.9424 \\
              & IMF\_3 & 0.0046 & 0.9978 & IMF\_3 & 0.0045 & 0.9960 \\
            \multirow{3}{*}{MSFT} 
              & IMF\_1 & 0.0130 & 0.9263 & IMF\_1 & 0.0547 & -2.9550 \\
              & IMF\_2 & 0.0072 & 0.9863 & IMF\_2 & 0.0184 & 0.8912 \\
              & IMF\_3 & 0.0088 & 0.9933 & IMF\_3 & 0.0104 & 0.9820 \\
            \multirow{3}{*}{GOOGL} 
              & IMF\_1 & 0.0181 & 0.8900 & IMF\_1 & 0.0397 & -0.3396 \\
              & IMF\_2 & 0.0081 & 0.9865 & IMF\_2 & 0.0130 & 0.9437 \\
              & IMF\_3 & 0.0081 & 0.9954 & IMF\_3 & 0.0075 & 0.9906 \\
            \multirow{3}{*}{NVDA} 
              & IMF\_1 & 0.0157 & 0.7815 & IMF\_1 & 0.0400 & -0.6051 \\
              & IMF\_2 & 0.0062 & 0.9851 & IMF\_2 & 0.0102 & 0.9434 \\
              & IMF\_3 & 0.0093 & 0.9909 & IMF\_3 & 0.0084 & 0.9874 \\
            \multirow{3}{*}{TSLA} 
              & IMF\_1 & 0.0218 & 0.8254 & IMF\_1 & 0.0310 & 0.5355 \\
              & IMF\_2 & 0.0066 & 0.9944 & IMF\_2 & 0.0125 & 0.9467 \\
              & IMF\_3 & 0.0105 & 0.9931 & IMF\_3 & 0.0070 & 0.9940 \\
            \multirow{3}{*}{AMGN} 
              & IMF\_1 & 0.0333 & 0.8620 & IMF\_1 & 0.0615 & 0.5971 \\
              & IMF\_2 & 0.0121 & 0.9927 & IMF\_2 & 0.0276 & 0.9450 \\
              & IMF\_3 & 0.0153 & 0.9934 & IMF\_3 & 0.0141 & 0.9917 \\
            \multirow{3}{*}{PEP} 
              & IMF\_1 & 0.0173 & 0.7539 & IMF\_1 & 0.0341 & 0.1095 \\
              & IMF\_2 & 0.0048 & 0.9882 & IMF\_2 & 0.0120 & 0.9430 \\
              & IMF\_3 & 0.0060 & 0.9968 & IMF\_3 & 0.0067 & 0.9907 \\
            \multirow{3}{*}{HON} 
              & IMF\_1 & 0.0235 & 0.8523 & IMF\_1 & 0.0457 & 0.3082 \\
              & IMF\_2 & 0.0088 & 0.9917 & IMF\_2 & 0.0129 & 0.9606 \\
              & IMF\_3 & 0.0109 & 0.9956 & IMF\_3 & 0.0067 & 0.9957 \\
            \bottomrule
        \end{tabular}
    \end{minipage}

\end{table*}

From the data in Table \ref{tab:results}, SSA-MAEMD-TCN is significantly better than MAEMD-TCN for all the stocks. For example, the RMSE of the former is 1.1355 in GooGL, while that of the latter is 1.7628, and the predicted values of SSA-MAEMD-TCN are closer to the actual values. The former also shows better performance in terms of MAPE and R². In addition, the prediction performance of SSA-MAEMD-TCN is better than that of all the benchmark models mentioned above. The advantage of the proposed model is the introduction of SSA for data preprocessing and noise reduction. This step dramatically reduces the impact of noise in financial data on subsequent predictions, enabling the model to accurately capture valid information in the data and improve the overall prediction results. Thus, the use of SSA for data preprocessing is necessary.

MAEMD-TCN outperforms MEMD-TCN in the prediction of most of the stocks. In the case of AAPL, for example, the RMSE of MAEMD-TCN is 1.9308, which is lower than that of MEMD-TCN (2.3602). Regarding MAPE, the value of MAEMD-TCN is 0.0089, which is lower than that of MEMD-TCN (0.0109), indicating that its prediction error is minor. R² also shows that the fit of the MAEMD-TCN model (0.9796) to the data is better than that of MEMD-TCN (0.9674). MA-EMD solves the problem of the deficiency of MEMD in multivariate decomposition. Secondly, it uses KLD to evaluate the frequency similarity of different variable IMFs to ensure the consistency and matching of the decomposed IMFs in frequency. This improved decomposition method provides TCN with higher-quality input features, enabling MAEMD-TCN to capture patterns and regularities in time series accurately.

In comparing MAEMD-LSTM and MAEMD-TCN, the later shows advantages in all three metrics. Compared with LSTM, the advantage of TCN is that it can effectively extend the sensory field to capture long-distance dependencies in time series, is less prone to the problem of gradient vanishing or gradient explosion during the training process, and has a more stable prediction performance. Therefore, MAEMD-TCN can more effectively extract the time series' short-range and long-range dependency features, thus realizing more accurate prediction results.

\begin{figure*}[htbp]
    \centering
    {\includegraphics[width=0.45\textwidth]{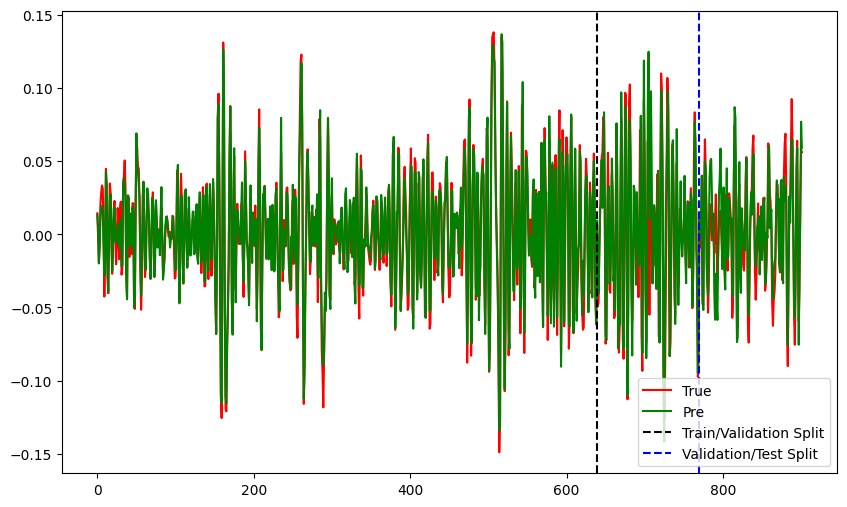}
    \hspace{0.02\textwidth} 
    {\includegraphics[width=0.45\textwidth]{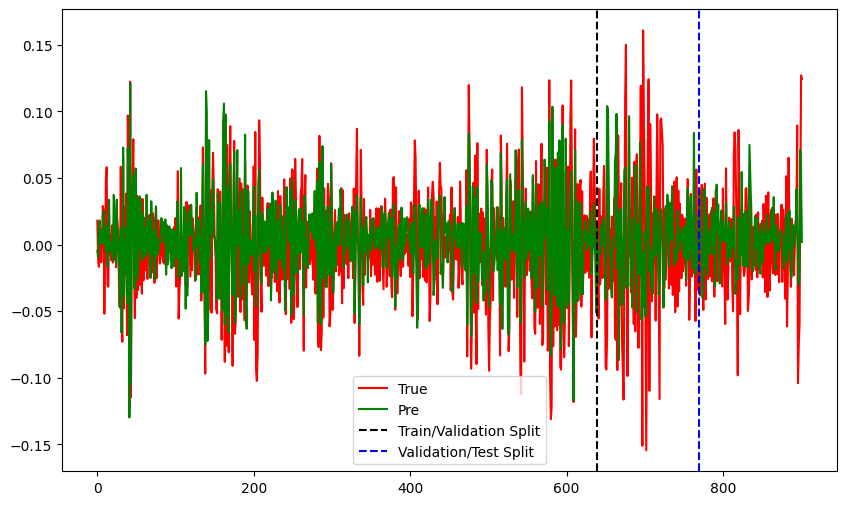}}
    \caption{Effectiveness of different models in fitting IMF\_1 of AAPL (The data has not been denormalized).}
    \label{fig:imf1}}
\end{figure*}

The following section will analyze how the noise reduction step improves the model performance. The EMD-based decomposition algorithm decomposes the sequence into multiple IMFs, and the forecasting model outputs the forecast values of these components and integrates them. The high-frequency components (such as IMF\_1) contain a lot of noise, which will interfere with the model's learning of effective trends. Through SSA denoising, SSA-MAEMD-TCN significantly outperforms MAEMD-TCN in predicting high-frequency components. As shown in Table \ref{tab:performance_comparison}, RMSE and R² indicators represent the model's learning ability. 

In IMF\_1 of AAPL, RMSE decreased from 0.036864 to 0.015491 (58.0\%), and MSFT from 0.054744 to 0.012974 (76.3\%). The R² indicator also improved significantly, from -0.339 to 0.890 in the GOOGL data. The denoising effect still exists in the sub-high frequency component (IMF\_2), but the improvement is weakened. For example, the RMSE of NVDA is reduced by 39.8\%, and the R² of GOOGLE is improved by 3.23\%. The noise impact is already low in the intermediate frequency component (IMF 3), and the denoising optimization effect is limited. In addition, combined with Fig. \ref{fig:imf1}, after denoising, the prediction model has better fitting effects on the training set, validation set, and test set.
These results show that SSA can effectively separate noise from trend information, allowing the model to more accurately capture the trend characteristics in the series.

\subsection{Portfolio Optimization}
In this section, we construct investor views based on the prediction results of the proposed model for eight stocks and construct an investment portfolio through the BL model. We select 132 trading days from February 23, 2023, to August 31, 2023 (out-of-sample data in the previous section) as the backtesting period. We select the mean-variance model (MV), the equal-weight model (EW), and the market-weighted portfolio model (MW) as benchmark models. This paper adopts two backtesting methods: "rolling data scheme" and "rebalancing strategy." In addition, we also compare the weight distribution of the portfolios constructed by the BL model and the MV model to evaluate whether the BL model can make up for the shortcomings of the MV model.

\subsubsection{Rolling data scheme}

This study uses a rolling data scheme to assess the effectiveness of portfolios under different holding periods. Starting from the initial date of the specified test period, the portfolio is constructed based on the forecast data of that day. It is assumed to remain unchanged for a predetermined holding period (e.g., 10 days). After the first holding period, the portfolio is reconstructed using forecast data from the second day of the test period, and the process is repeated until the end of the test period. The portfolio's overall performance is assessed by calculating and annualizing the average return over the holding period, which is taken to be 1, 3, 5, 10, and 20 days. The focus is on observing the relationship between annualized returns and Sharpe ratios and the length of the holding period.
\begin{figure*}[htbp]
    \centering
    {\includegraphics[width=0.45\textwidth]{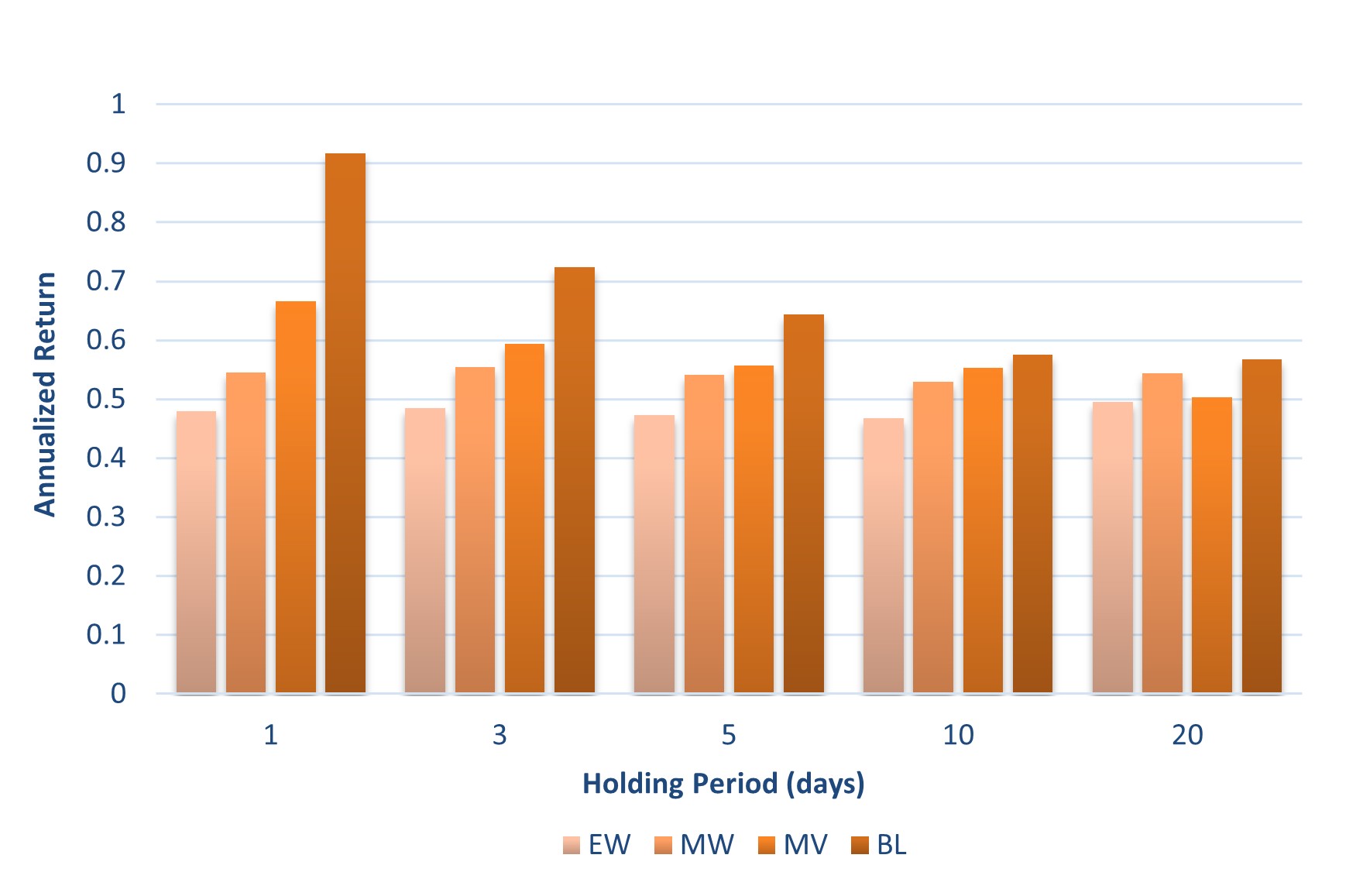}}\label{fig:rolling left}
    \hspace{0.02\textwidth} 
    {\includegraphics[width=0.45\textwidth]{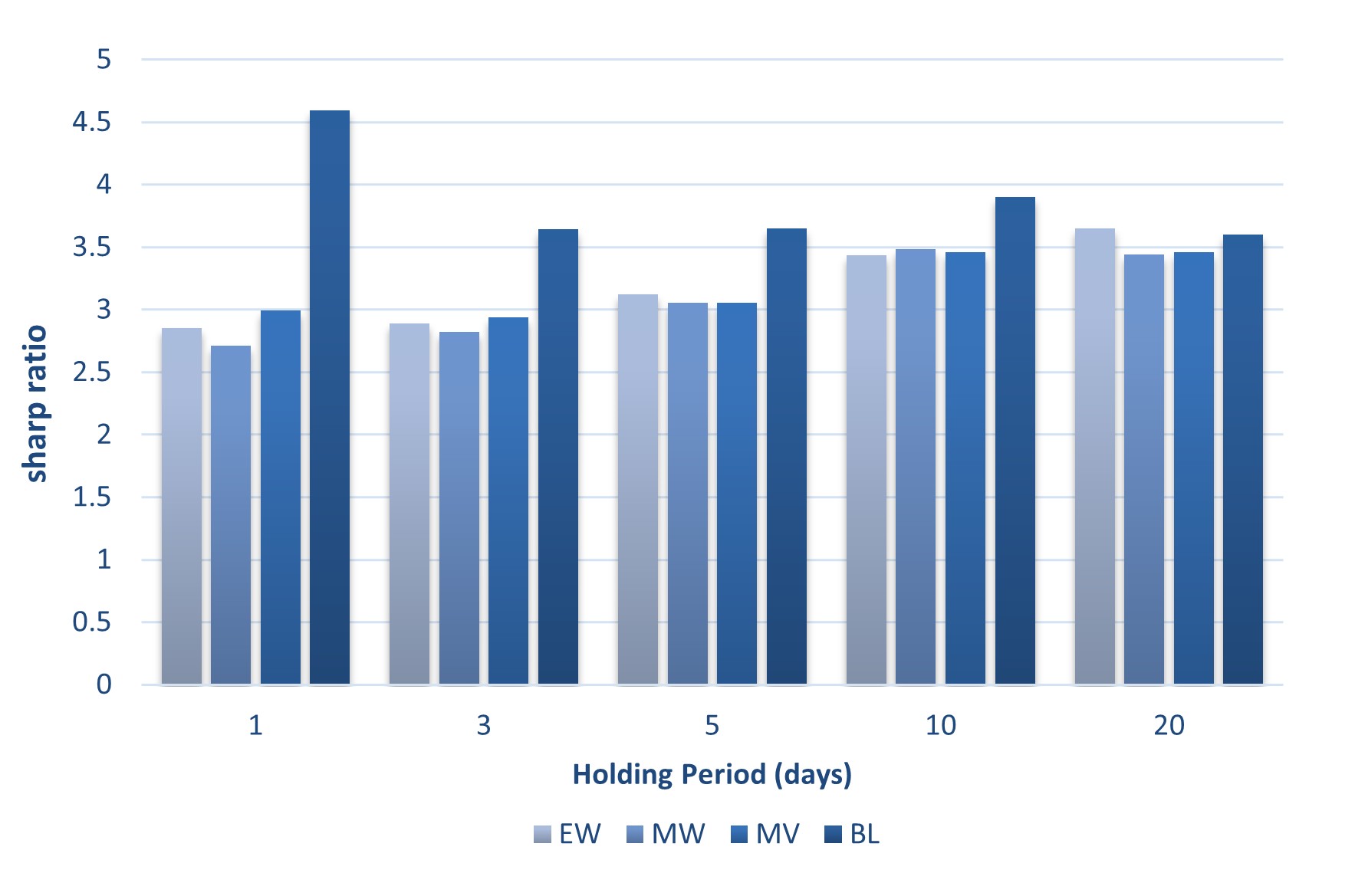}}
    \label{fig:rolling right}
    \caption{Evaluate portfolio performance using annualized returns and Sharpe ratios.}
    \label{fig:rolling window}
\end{figure*}

The empirical results in Fig. \ref{fig:rolling window} confirm the significant advantages of the BL model for short holding periods. When looking at the overall data, all strategies exhibit high Sharpe ratios and decent annualized returns under different holding periods, indicating that the market is in a state where the trend is more pronounced or the volatility is relatively orderly, giving all investment strategies a chance to achieve profitability.

\begin{table*}[ht]
\centering
\caption{The number of times with the highest and lowest return at rolling runs.}
\Large
\resizebox{\textwidth}{!}{
\begin{tabular}{cccccccccc}
\toprule
\multirow{2}{*}{Period} & \multirow{2}{*}{Number of Runs} & \multicolumn{4}{c}{Highest Return} & \multicolumn{4}{c}{Lowest Return} \\
\cmidrule(lr){3-6} \cmidrule(lr){7-10}
& & BL-Highest & EW-Highest & MW-Highest & MV-Highest & BL-Lowest & EW-Lowest & MW-Lowest & MV-Lowest \\
\midrule
1 & 132 & 53 (40.15\%) & 24 & 1 & 54 & 7 (5.30\%) & 22 & 49 & 54 \\
3 & 130 & 55 (42.31\%) & 20 & 10 & 45 & 7 (5.38\%) & 40 & 32 & 51 \\
5 & 128 & 40 (31.25\%) & 19 & 14 & 55 & 14 (10.94\%)& 32 & 31 & 51 \\
10 & 123 & 28 (22.76\%) & 25 & 22 & 48 & 8 (6.50\%)& 46 & 24 & 45 \\
20 & 113 & 26 (23.01\%) & 25 & 24 & 38 & 14 (12.39\%)& 29 & 26 & 44 \\
\midrule
Overall & 626 & 202 & 113 & 71 & 240 & 50 & 169 & 162 & 245 \\
\midrule
Percentage & & 32.27\% & 18.05\% & 11.34\% & 38.34\% & 7.99\% &  27.00\%& 25.88\% & 39.14\% \\
\bottomrule
\label{tab:Performance}
\end{tabular}}
\end{table*}

With a holding period of 1 day, the BL model achieves an annualized return of more than 80\% and a Sharpe ratio of about 4.5, which significantly outperforms the other models. This is mainly due to the proposed hybrid model, which improves the accuracy of the subjective view. By keenly capturing market fluctuations, the BL model adequately adjusts the portfolio, resulting in high returns and better risk control during the short holding period. As the holding period lengthens, the advantages of BL's annualized returns and Sharpe ratios diminish. Longer periods increase market uncertainty, and subjective views are no longer adapted to market conditions. Based on these outdated views, portfolio adjustments are no longer effective, and yield and volatility control advantages are no longer as pronounced in the short term.

To further verify the above experimental findings, we count the times the four models obtain the highest return and the times they get the lowest return under different holding periods, as shown in Table \ref{tab:Performance}. For the 1-day and 3-day holding periods, the BL model obtains the highest return 53 times and 55 times, accounting for 40.15\% and 42.31\%, respectively. In contrast, the lowest returns are only 7 times, 5.30\% and 5.38\%, highlighting its stability and superiority in short-term investment. When the holding period is gradually extended, the number of times that the BL model obtains the highest rate of return gradually decreases, and the number of times that it obtains the lowest rate of return gradually increases, especially when the holding period is 20 days, it reaches 14 times, accounting for 12.39\%, which indicates that its volatility and riskiness increase in a more extended holding period.

\begin{figure*}[h]
    \centering
    \includegraphics[width=0.8\textwidth]{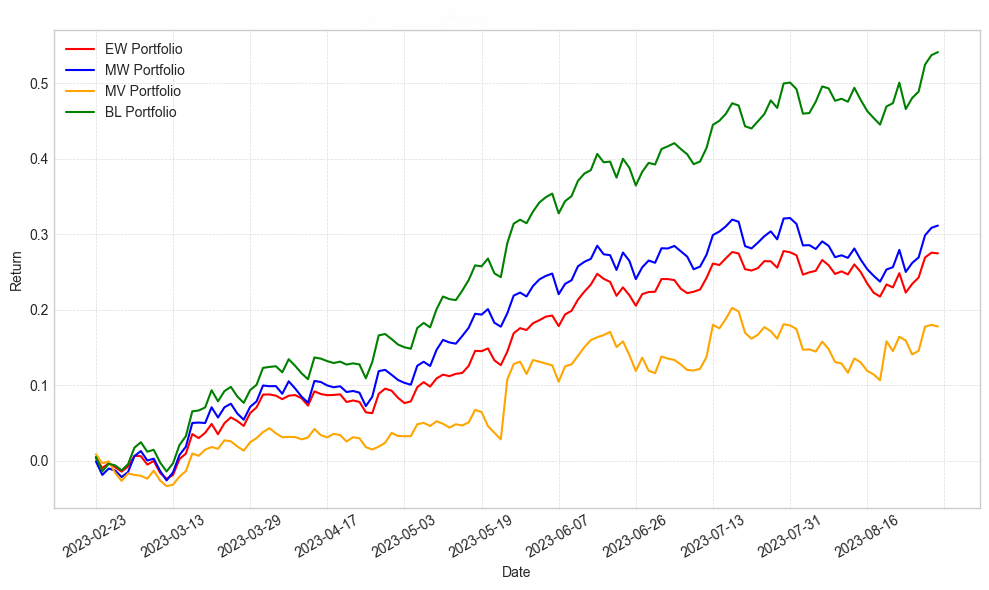}
    \caption{Cumulative return of 1-day rebalancing.}
    \label{fig:1dcumulative}
\end{figure*}

\subsubsection{Rebalancing Strategy}
The rebalancing strategy dynamically manages an investment portfolio by adjusting asset weights to meet the investor's initial goals. When market fluctuations cause asset weights to deviate from the preset ratio, the target weights are restored by buying and selling assets. This adjustment can be a regular adjustment based on time intervals or an irregular adjustment based on the degree of deviation of asset weights. This paper uses the "periodic rebalancing strategy" for experiments and assumes that the transaction cost is 0.2\%. The holding period is consistent with the previous article. Table~\ref{tab:rebalance} shows the performance of four different portfolio strategies under different holding periods.

\begin{table}[ht]
\centering
\setlength{\tabcolsep}{3pt} 
\caption{Performance of portfolio management with rebalancing }

\begin{tabular}{@{}cccccc@{}}
\toprule
Period & Metrics& BL & MV & EW & MW \\
\toprule
\multirow{3}{*}{1} &  Annual Return   &1.0335 & 0.3399  & 0.5247 & 0.5951 \\
 & Annual Volatility  & 0.1889  &0.2038  &  0.1508 & 0.1826 \\
 & Sharpe Ratio &  4.2154 & 1.3967 & 2.8288 &  2.6629\\

\multirow{3}{*}{3} &  Annual Return  &0.7963 & 0.5411  & 0.5240 & 0.5973 \\
 & Annual Volatility  & 0.1748  &0.1841  &  0.1437 & 0.1731\\
 & Sharpe Ratio &  3.6301 & 2.4209  & 2.9633 & 2.8157 \\

\multirow{3}{*}{5} &  Annual Return  &0.6689 & 0.5801  & 0.5236 & 0.6010 \\
 & Annual Volatility  & 0.1660  &0.1466  &  0.1370 & 0.1647\\
 & Sharpe Ratio &  3.1850 & 3.1359  & 3.0292 & 2.9027 \\

\multirow{3}{*}{10} &  Annual Return  &0.6056 & 0.5750  & 0.5261 & 0.5996\\
 & Annual Volatility  & 0.1317  & 0.1332  &  0.1174 & 0.1390\\
 & Sharpe Ratio & 3.4953  &3.2902  & 3.4048 &  3.2879\\

 \multirow{3}{*}{20} &  Annual Return & 0.6264 & 0.5322  & 0.5368 & 0.5946 \\
 & Annual Volatility  & 0.1424  & 0.1604 &  0.1095 &  0.1386\\
 & Sharpe Ratio &3.3877  & 2.5836  & 3.7542 &   3.3066\\

\bottomrule
\end{tabular}
\label{tab:rebalance}

\end{table}

The BL model shows a significant holding period effect: when the holding period is extended from 1 day to 20 days, the expected annualized return decreases from 1.0335 to 0.6264, and the Sharpe ratio decreases from 4.2154 to 3.3877. It is the only strategy whose Sharpe ratio decreases with the increase of the holding period. This shows that the BL model achieves higher returns in a short holding period by combining market equilibrium returns and subjective views through a Bayesian framework. Although the transaction costs are high, its excellent performance makes the strategy returns less affected by transaction costs. However, as the holding period increases, the adaptability of investors' views decreases, and the performance of the BL model gradually approaches that of the MW model.
The annualized return of the MV model shows a trend of first rising and then falling with the increase of the holding period: from 0.3399 in a 1-day holding period to 0.5801 in a 5-day holding period, and then falls back to 0.5322 in 20 days (a decrease of 7.4\%), while the volatility shows a trend of first falling and then rising. In the short holding period, the Sharpe ratio of the MV model is significantly lower than that of the BL model, exposing the problems of high transaction costs and parameter misalignment under high-frequency adjustments. 

In contrast, the EW and MW portfolios maintain relatively stable annualized returns under different holding periods, with gradually decreasing volatility and gradually increasing Sharpe ratios, verifying the characteristics of passive allocation to achieve long-term stability by diversifying risks. However, the passive investment strategy lacks flexibility, and its returns and Sharpe ratios are lower than the BL model's in the short holding period (1-3 days).

\begin{table}[ht]

\caption{ Performance of portfolio management with 1-day rebalancing}
\setlength{\tabcolsep}{2pt} 
\centering

\begin{tabular}{lcccc}
\toprule
& EW & MW & MV & BL \\
\midrule
daily average return (\%)  & 0.1886 & 0.2124  & 0.1323  &0.3354  \\
daily volatility(\%)  & 0.9498 & 1.1505 & 1.2840  & 1.1900  \\
Cumulative Return & 0.2748 & 0.3117 & 0.1780 & 0.5414 \\
Sharpe Ratio & 2.8288 & 2.6629 & 1.3967 &  4.2154 \\ 
\bottomrule
\label{tab:1drebalance}
\end{tabular}

\end{table}

\begin{figure*}[htbp]
    \centering
        {\includegraphics[width=0.49\textwidth]{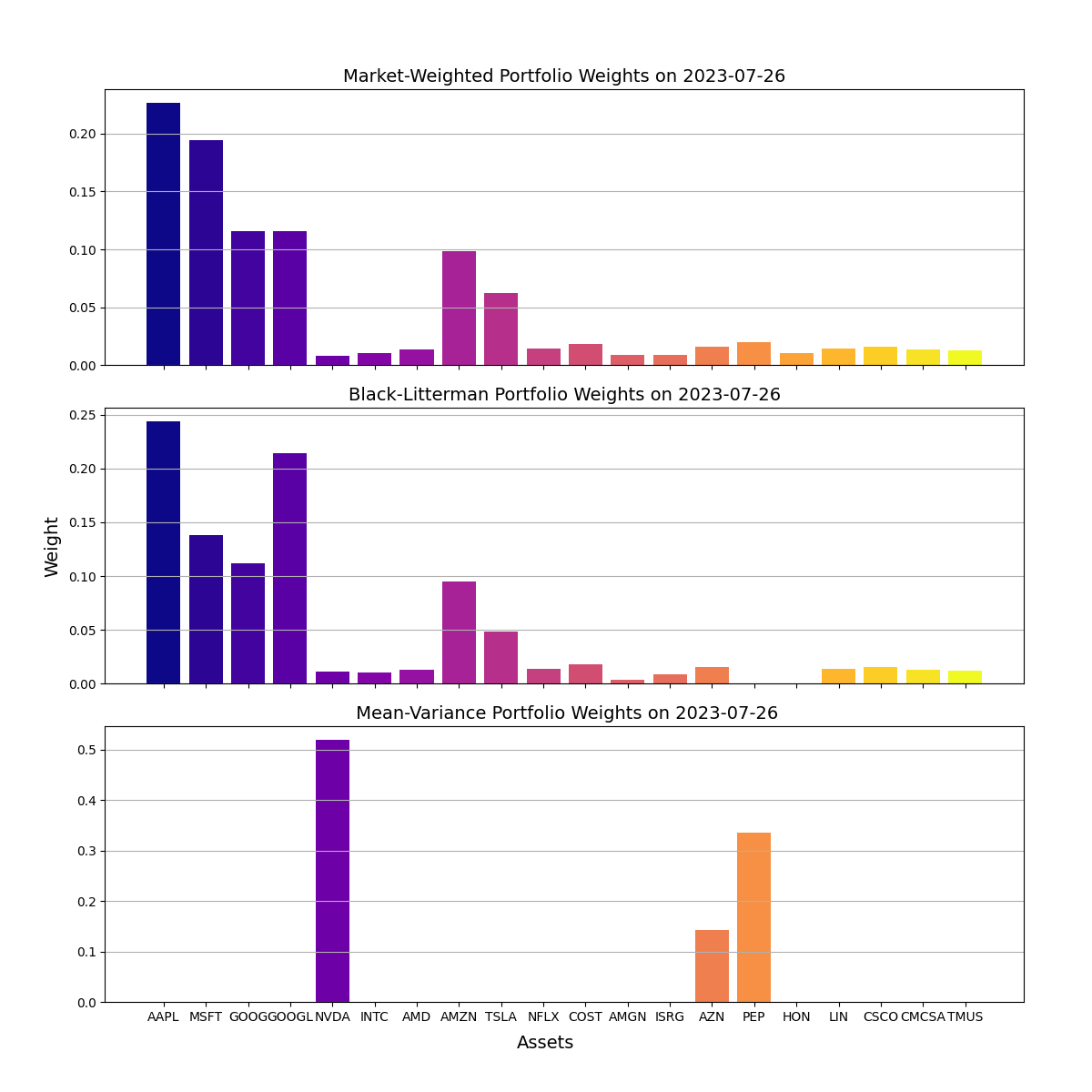}}
    \hfill
        \centering
        \raisebox{2cm}{\includegraphics[width=0.49\textwidth]{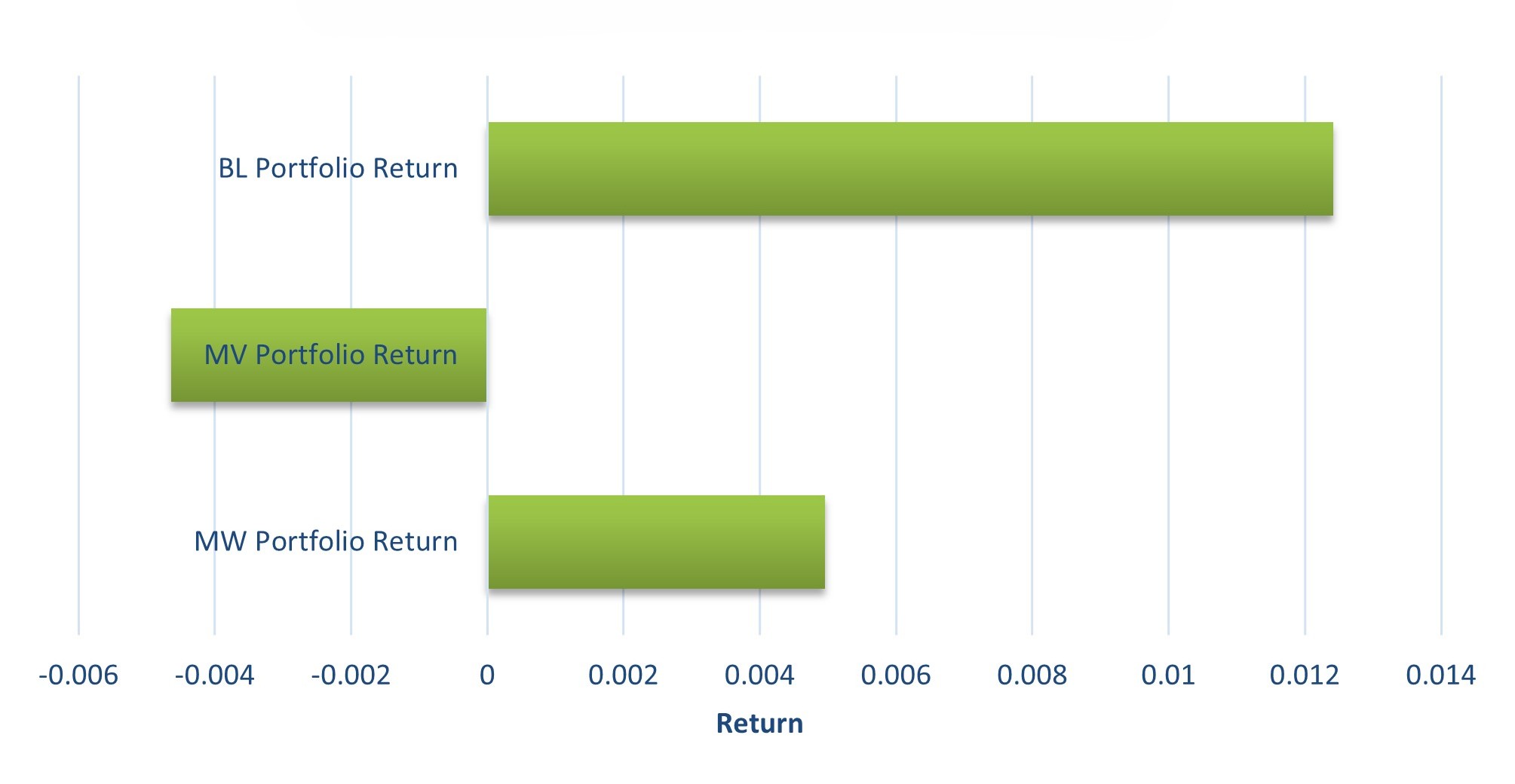}}
    \caption{Comparison of different portfolio weights and returns in 2023.07.26}
    \label{fig:adjusting}
\end{figure*}
We focus on analyzing a special case of the rebalancing strategy: 1-day rebalancing, which is essentially a portfolio model that generates daily portfolio weights, as shown in Table \ref{tab:1drebalance} and Fig. \ref{fig:1dcumulative}. In the framework of this strategy, the BL model has a significantly higher average daily and cumulative return than the other strategies, with a Sharpe ratio of 4.2154, much higher than the MV model's 1.3967. Although the daily volatility of 1.190\% is slightly higher than that of the equal-weighted portfolio's 0.9498\%, the high Sharpe ratio confirms the effectiveness of this strategy dynamically adjusting to the short holding period. Combined with the accurate prediction results of the hybrid model, the BL model achieves a balance between maximizing return and controlling risk in a very short holding period.

The EW portfolio reflects the defensive nature of the passive strategy with the lowest volatility and stable returns. Still, its cumulative return is only 50\% of the BL model's. The MV model has the lowest daily return and the highest daily volatility. These data show that its reliance on historical data for optimization has the risk of overfitting, and the transaction costs of high-frequency position adjustments further weaken the model performance. In short, the BL model is thoroughly dominant in high-frequency active investment strategies.

\subsubsection{Portfolio Diversification}

In this section, we discuss the degree of portfolio diversification for the Black-Litterman and Mean-Variance models. We select the daily weights under 1-day rebalance to test whether the BL model has an advantage in terms of the degree of diversification of the underlying. As seen from the two-dimensional empirical results of weight distribution and Herfindahl Index (Table \ref{tab:concentration}), the Black-Litterman and the Mean-Variance models present a significant difference in underlying diversification.
In terms of weight distribution, the BL model covers 18.03 stocks with a standard deviation of 1.039, which is significantly better than the MV model's 11.19 stocks. Meanwhile, the former's standard deviation is even lower than that of the latter, suggesting that the BL model has a more stable underlying distribution. 
\begin{table}[ht]
\caption{Portfolio Weight Distribution}
\centering

\begin{tabular}{lcccc}
\toprule
& MV & BL&\\
\midrule
Mean of stock number & 11.190 &18.030    \\
Std of stock number  &8.183&    0.937    \\
Mean of HHI & 0.245 & 0.131\\
Std of HHI & 0.157 &  0.009\\
\bottomrule
\end{tabular}
\label{tab:concentration}
\end{table}

Further deconstructing the concentration difference by HHI, the mean HHI of the BL model is 0.131, which is much lower than that of the MV model (0.245). Its HHI standard deviation is only 0.009, which is equivalent to 5.73\% of the volatility of the MV model. The MV model presents high concentration and allocation instability compared to the BL model. Regarding portfolio diversification, the BL model is superior across the board in the diversification and robustness dimensions.

Fig.~\ref{fig:adjusting} shows the weights and returns of different investment portfolios on a particular day. The BL model combines the views of investors, so its investment portfolio can be regarded as an optimization of the MW portfolio. Therefore, ideally, the single-day return of the BL model is higher. The portfolio distribution and return of the MV model in this day are not good enough.



\section{Conclusions}
The traditional Black-Litterman model faces challenges in the generation of investor views. This study innovatively constructs an SSA-MAEMD-TCN hybrid deep learning framework, which integrates Singular Spectrum Analysis (SSA), Multivariate Aligned Empirical Mode Decomposition (MA-EMD), and Temporal Convolutional Networks (TCN) to achieve high-precision stock price forecasting and provide reliable inputs for the Black-Litterman model. Empirical results demonstrate that, in terms of asset price prediction, the hybrid model significantly outperforms three benchmark models—MAEMD-TCN, MEMD-TCN, and MAEMD-LSTM—in key metrics such as RMSE, MAPE, and R². In addition, we also find that the prediction performance of medium and high frequency sequences can be significantly improved by using the SSA algorithm to process the noise reduction of the original sequences

In portfolio management, we conducted a comprehensive performance evaluation of the investment portfolios constructed based on this model using two backtesting methods: the rolling data scheme and the rebalancing strategy. Comparative analyses with benchmark models such as the Mean-Variance, Equal-Weighted, and Market-Weighted models further illustrate the superiority of the investment portfolios generated by combining the hybrid and Black-Litterman models. Additionally, we examined the diversification of the investment portfolios. Compared with the mean-variance model, the Black-Litterman model can generate more diversified portfolios, effectively reducing concentration risk and enhancing portfolio robustness.

Future research can focus on two main directions. First, the current model primarily relies on daily data for single-step predictions, which is suitable for short-term trading but has limitations in medium- and long-term investments. By developing multi-step prediction models, we can forecast stock price trends over longer horizons, such as one week, one month, or even longer. This enables investors to make informed decisions for medium- and long-term investments. Second, further research is needed to improve portfolio optimization methods, especially in dynamic environments. Techniques such as reinforcement learning and online learning can be explored to develop adaptive portfolio optimization strategies that allow investment portfolios to automatically adjust and optimize in response to changing market conditions.

\bibliography{sn-bibliography}

\begin{thebibliography}{}
\renewcommand{\doi}[1]{\url{https://doi.org/#1}}
\bibcommenthead

\bibitem [\protect \citeauthoryear {%
Ak{\c{s}}ehir%
\ \BBA {} K{\i}l{\i}{\c{c}}%
}{%
Ak{\c{s}}ehir%
\ \BBA {} K{\i}l{\i}{\c{c}}%
}{%
{\protect \APACyear {2024}}%
}]{%
aksehirNewDenoisingApproach2024a}
\APACinsertmetastar {%
aksehirNewDenoisingApproach2024a}%
\begin{APACrefauthors}%
Ak{\c{s}}ehir, Z.D.%
\BCBT {}\ \BBA {} K{\i}l{\i}{\c{c}}, E.%
\end{APACrefauthors}%
\unskip\
\newblock
\APACrefYearMonthDay{2024}{}{}.
\newblock
{\BBOQ}\APACrefatitle {A new denoising approach based on mode decomposition applied to the stock market time series: 2LE-CEEMDAN} {A new denoising approach based on mode decomposition applied to the stock market time series: 2le-ceemdan}.{\BBCQ}
\newblock
\APACjournalVolNumPages{PeerJ Computer Science}{10}{}{e1852,}
\newblock
\begin{APACrefDOI} \doi{10.7717/peerj-cs.1852} \end{APACrefDOI}
\newblock

\newblock

\PrintBackRefs{\CurrentBib}

\bibitem [\protect \citeauthoryear {%
Alrumaih%
\ \BBA {} Al-Fawzan%
}{%
Alrumaih%
\ \BBA {} Al-Fawzan%
}{%
{\protect \APACyear {2002}}%
}]{%
alrumaihTimeSeriesForecasting2002}
\APACinsertmetastar {%
alrumaihTimeSeriesForecasting2002}%
\begin{APACrefauthors}%
Alrumaih, R.M.%
\BCBT {}\ \BBA {} Al-Fawzan, M.A.%
\end{APACrefauthors}%
\unskip\
\newblock
\APACrefYearMonthDay{2002}{}{}.
\newblock
{\BBOQ}\APACrefatitle {Time series forecasting using wavelet denoising an application to saudi stock index} {Time series forecasting using wavelet denoising an application to saudi stock index}.{\BBCQ}
\newblock
\APACjournalVolNumPages{Journal of King Saud University-Engineering Sciences}{14}{2}{221--233,}
\newblock
\begin{APACrefDOI} \doi{10.1016/S1018-3639(18)30755-4} \end{APACrefDOI}
\newblock

\newblock

\PrintBackRefs{\CurrentBib}

\bibitem [\protect \citeauthoryear {%
J.~Bai%
\ \protect \BOthers {.}}{%
J.~Bai%
\ \protect \BOthers {.}}{%
{\protect \APACyear {2023}}%
}]{%
baiIntelligentForecastingModel2023}
\APACinsertmetastar {%
baiIntelligentForecastingModel2023}%
\begin{APACrefauthors}%
Bai, J.%
, Guo, J.%
, Sun, B.%
, Guo, Y.%
, Bao, Q.%
\BCBL {} Xiao, X.%
\end{APACrefauthors}%
\unskip\
\newblock
\APACrefYearMonthDay{2023}{}{}.
\newblock
{\BBOQ}\APACrefatitle {Intelligent forecasting model of stock price using neighborhood rough set and multivariate empirical mode decomposition} {Intelligent forecasting model of stock price using neighborhood rough set and multivariate empirical mode decomposition}.{\BBCQ}
\newblock
\APACjournalVolNumPages{Engineering Applications of Artificial Intelligence}{122}{}{106106,}
\newblock
\begin{APACrefDOI} \doi{10.1016/j.engappai.2023.106106} \end{APACrefDOI}
\newblock

\newblock

\PrintBackRefs{\CurrentBib}

\bibitem [\protect \citeauthoryear {%
S.~Bai%
, Kolter%
\BCBL {}\ \BBA {} Koltun%
}{%
S.~Bai%
\ \protect \BOthers {.}}{%
{\protect \APACyear {2018}}%
}]{%
baiEmpiricalEvaluationGeneric2018}
\APACinsertmetastar {%
baiEmpiricalEvaluationGeneric2018}%
\begin{APACrefauthors}%
Bai, S.%
, Kolter, J.Z.%
\BCBL {} Koltun, V.%
\end{APACrefauthors}%
\unskip\
\newblock
\APACrefYearMonthDay{2018}{{\APACmonth{04}}}{}.
\newblock
\APACrefbtitle {An {{Empirical Evaluation}} of {{Generic Convolutional}} and {{Recurrent Networks}} for {{Sequence Modeling}}} {An {{Empirical Evaluation}} of {{Generic Convolutional}} and {{Recurrent Networks}} for {{Sequence Modeling}}}\ (\BNUM\ arXiv:1803.01271).
\newblock
\APACaddressPublisher{}{arXiv}.
\PrintBackRefs{\CurrentBib}

\bibitem [\protect \citeauthoryear {%
Bao%
, Yue%
\BCBL {}\ \BBA {} Rao%
}{%
Bao%
\ \protect \BOthers {.}}{%
{\protect \APACyear {2017}}%
}]{%
baoDeepLearningFramework2017}
\APACinsertmetastar {%
baoDeepLearningFramework2017}%
\begin{APACrefauthors}%
Bao, W.%
, Yue, J.%
\BCBL {} Rao, Y.%
\end{APACrefauthors}%
\unskip\
\newblock
\APACrefYearMonthDay{2017}{}{}.
\newblock
{\BBOQ}\APACrefatitle {A deep learning framework for financial time series using stacked autoencoders and long-short term memory} {A deep learning framework for financial time series using stacked autoencoders and long-short term memory}.{\BBCQ}
\newblock
\APACjournalVolNumPages{PloS one}{12}{7}{e0180944,}
\newblock
\begin{APACrefDOI} \doi{10.1371/journal.pone.0180944} \end{APACrefDOI}
\newblock

\newblock

\PrintBackRefs{\CurrentBib}

\bibitem [\protect \citeauthoryear {%
Barua%
\ \BBA {} Sharma%
}{%
Barua%
\ \BBA {} Sharma%
}{%
{\protect \APACyear {2022}}%
}]{%
baruaDynamicBlackLitterman2022}
\APACinsertmetastar {%
baruaDynamicBlackLitterman2022}%
\begin{APACrefauthors}%
Barua, R.%
\BCBT {}\ \BBA {} Sharma, A.K.%
\end{APACrefauthors}%
\unskip\
\newblock
\APACrefYearMonthDay{2022}{}{}.
\newblock
{\BBOQ}\APACrefatitle {Dynamic Black Litterman portfolios with views derived via CNN-BiLSTM predictions} {Dynamic black litterman portfolios with views derived via cnn-bilstm predictions}.{\BBCQ}
\newblock
\APACjournalVolNumPages{Finance Research Letters}{49}{}{103111,}
\newblock
\begin{APACrefDOI} \doi{10.1016/j.frl.2022.103111} \end{APACrefDOI}
\newblock

\newblock

\PrintBackRefs{\CurrentBib}

\bibitem [\protect \citeauthoryear {%
Barua%
\ \BBA {} Sharma%
}{%
Barua%
\ \BBA {} Sharma%
}{%
{\protect \APACyear {2023}}%
}]{%
baruaUsingFearGreed2023}
\APACinsertmetastar {%
baruaUsingFearGreed2023}%
\begin{APACrefauthors}%
Barua, R.%
\BCBT {}\ \BBA {} Sharma, A.K.%
\end{APACrefauthors}%
\unskip\
\newblock
\APACrefYearMonthDay{2023}{}{}.
\newblock
{\BBOQ}\APACrefatitle {Using fear, greed and machine learning for optimizing global portfolios: A Black-Litterman approach} {Using fear, greed and machine learning for optimizing global portfolios: A black-litterman approach}.{\BBCQ}
\newblock
\APACjournalVolNumPages{Finance Research Letters}{58}{}{104515,}
\newblock
\begin{APACrefDOI} \doi{10.1016/j.frl.2023.104515} \end{APACrefDOI}
\newblock

\newblock

\PrintBackRefs{\CurrentBib}

\bibitem [\protect \citeauthoryear {%
Beach%
\ \BBA {} Orlov%
}{%
Beach%
\ \BBA {} Orlov%
}{%
{\protect \APACyear {2007}}%
}]{%
beachApplicationBlackLitterman2007a}
\APACinsertmetastar {%
beachApplicationBlackLitterman2007a}%
\begin{APACrefauthors}%
Beach, S.L.%
\BCBT {}\ \BBA {} Orlov, A.G.%
\end{APACrefauthors}%
\unskip\
\newblock
\APACrefYearMonthDay{2007}{}{}.
\newblock
{\BBOQ}\APACrefatitle {An application of the Black--Litterman model with EGARCH-M-derived views for international portfolio management} {An application of the black--litterman model with egarch-m-derived views for international portfolio management}.{\BBCQ}
\newblock
\APACjournalVolNumPages{Financial Markets and Portfolio Management}{21}{}{147--166,}
\newblock
\begin{APACrefDOI} \doi{10.1007/s11408-007-0046-6} \end{APACrefDOI}
\newblock

\newblock

\PrintBackRefs{\CurrentBib}

\bibitem [\protect \citeauthoryear {%
Best%
\ \BBA {} Grauer%
}{%
Best%
\ \BBA {} Grauer%
}{%
{\protect \APACyear {1991}}%
}]{%
bestSensitivityMeanVarianceEfficientPortfolios1991}
\APACinsertmetastar {%
bestSensitivityMeanVarianceEfficientPortfolios1991}%
\begin{APACrefauthors}%
Best, M.J.%
\BCBT {}\ \BBA {} Grauer, R.R.%
\end{APACrefauthors}%
\unskip\
\newblock
\APACrefYearMonthDay{1991}{}{}.
\newblock
{\BBOQ}\APACrefatitle {On the sensitivity of mean-variance-efficient portfolios to changes in asset means: some analytical and computational results} {On the sensitivity of mean-variance-efficient portfolios to changes in asset means: some analytical and computational results}.{\BBCQ}
\newblock
\APACjournalVolNumPages{The review of financial studies}{4}{2}{315--342,}
\newblock
\begin{APACrefDOI} \doi{10.1093/rfs/4.2.315} \end{APACrefDOI}
\newblock

\newblock

\PrintBackRefs{\CurrentBib}

\bibitem [\protect \citeauthoryear {%
Black%
}{%
Black%
}{%
{\protect \APACyear {1986}}%
}]{%
blackNoise1986}
\APACinsertmetastar {%
blackNoise1986}%
\begin{APACrefauthors}%
Black, F.%
\end{APACrefauthors}%
\unskip\
\newblock
\APACrefYearMonthDay{1986}{}{}.
\newblock
{\BBOQ}\APACrefatitle {Noise} {Noise}.{\BBCQ}
\newblock
\APACjournalVolNumPages{The journal of finance}{41}{3}{528--543,}
\newblock
\begin{APACrefDOI} \doi{10.1111/j.1540-6261.1986.tb04513.x} \end{APACrefDOI}
\newblock

\newblock

\PrintBackRefs{\CurrentBib}

\bibitem [\protect \citeauthoryear {%
Black%
\ \BBA {} Litterman%
}{%
Black%
\ \BBA {} Litterman%
}{%
{\protect \APACyear {1992}}%
}]{%
blackGlobalPortfolioOptimization1992}
\APACinsertmetastar {%
blackGlobalPortfolioOptimization1992}%
\begin{APACrefauthors}%
Black, F.%
\BCBT {}\ \BBA {} Litterman, R.%
\end{APACrefauthors}%
\unskip\
\newblock
\APACrefYearMonthDay{1992}{}{}.
\newblock
{\BBOQ}\APACrefatitle {Global portfolio optimization} {Global portfolio optimization}.{\BBCQ}
\newblock
\APACjournalVolNumPages{Financial analysts journal}{48}{5}{28--43,}
\newblock
\begin{APACrefDOI} \doi{10.2469/faj.v48.n5.28} \end{APACrefDOI}
\newblock

\newblock

\PrintBackRefs{\CurrentBib}

\bibitem [\protect \citeauthoryear {%
Broomhead%
\ \BBA {} King%
}{%
Broomhead%
\ \BBA {} King%
}{%
{\protect \APACyear {1986}}%
}]{%
broomheadExtractingQualitativeDynamics1986}
\APACinsertmetastar {%
broomheadExtractingQualitativeDynamics1986}%
\begin{APACrefauthors}%
Broomhead, D.S.%
\BCBT {}\ \BBA {} King, G.P.%
\end{APACrefauthors}%
\unskip\
\newblock
\APACrefYearMonthDay{1986}{}{}.
\newblock
{\BBOQ}\APACrefatitle {Extracting qualitative dynamics from experimental data} {Extracting qualitative dynamics from experimental data}.{\BBCQ}
\newblock
\APACjournalVolNumPages{Physica D: Nonlinear Phenomena}{20}{2-3}{217--236,}
\newblock
\begin{APACrefDOI} \doi{10.1016/0167-2789(86)90031-X} \end{APACrefDOI}
\newblock

\newblock

\PrintBackRefs{\CurrentBib}

\bibitem [\protect \citeauthoryear {%
Cai%
\ \BBA {} Li%
}{%
Cai%
\ \BBA {} Li%
}{%
{\protect \APACyear {2024}}%
}]{%
caiMEDEMMNNbasedEmpirical2024}
\APACinsertmetastar {%
caiMEDEMMNNbasedEmpirical2024}%
\begin{APACrefauthors}%
Cai, X.%
\BCBT {}\ \BBA {} Li, D.%
\end{APACrefauthors}%
\unskip\
\newblock
\APACrefYearMonthDay{2024}{}{}.
\newblock
{\BBOQ}\APACrefatitle {M-EDEM: A MNN-based Empirical Decomposition Ensemble Method for improved time series forecasting} {M-edem: A mnn-based empirical decomposition ensemble method for improved time series forecasting}.{\BBCQ}
\newblock
\APACjournalVolNumPages{Knowledge-Based Systems}{283}{}{111157,}
\newblock
\begin{APACrefDOI} \doi{10.1016/j.knosys.2023.111157} \end{APACrefDOI}
\newblock

\newblock

\PrintBackRefs{\CurrentBib}

\bibitem [\protect \citeauthoryear {%
Cai%
, Li%
, Zhang%
\BCBL {}\ \BBA {} Wu%
}{%
Cai%
\ \protect \BOthers {.}}{%
{\protect \APACyear {2025}}%
}]{%
caiMAEMDAlignedEmpirical2025}
\APACinsertmetastar {%
caiMAEMDAlignedEmpirical2025}%
\begin{APACrefauthors}%
Cai, X.%
, Li, D.%
, Zhang, J.%
\BCBL {} Wu, Z.%
\end{APACrefauthors}%
\unskip\
\newblock
\APACrefYearMonthDay{2025}{}{}.
\newblock
{\BBOQ}\APACrefatitle {MA-EMD: Aligned empirical decomposition for multivariate time-series forecasting} {Ma-emd: Aligned empirical decomposition for multivariate time-series forecasting}.{\BBCQ}
\newblock
\APACjournalVolNumPages{Expert Systems with Applications}{267}{}{126080,}
\newblock
\begin{APACrefDOI} \doi{10.1016/j.eswa.2024.126080} \end{APACrefDOI}
\newblock

\newblock

\PrintBackRefs{\CurrentBib}

\bibitem [\protect \citeauthoryear {%
Deng%
, Huang%
, Hasan%
\BCBL {}\ \BBA {} Bao%
}{%
Deng%
\ \protect \BOthers {.}}{%
{\protect \APACyear {2022}}%
}]{%
dengMultistepaheadStockPrice2022}
\APACinsertmetastar {%
dengMultistepaheadStockPrice2022}%
\begin{APACrefauthors}%
Deng, C.%
, Huang, Y.%
, Hasan, N.%
\BCBL {} Bao, Y.%
\end{APACrefauthors}%
\unskip\
\newblock
\APACrefYearMonthDay{2022}{}{}.
\newblock
{\BBOQ}\APACrefatitle {Multi-step-ahead stock price index forecasting using long short-term memory model with multivariate empirical mode decomposition} {Multi-step-ahead stock price index forecasting using long short-term memory model with multivariate empirical mode decomposition}.{\BBCQ}
\newblock
\APACjournalVolNumPages{Information Sciences}{607}{}{297--321,}
\newblock
\begin{APACrefDOI} \doi{10.1016/j.ins.2022.05.088} \end{APACrefDOI}
\newblock

\newblock

\PrintBackRefs{\CurrentBib}

\bibitem [\protect \citeauthoryear {%
Duqi%
, Franci%
\BCBL {}\ \BBA {} Torluccio%
}{%
Duqi%
\ \protect \BOthers {.}}{%
{\protect \APACyear {2014}}%
}]{%
duqiBlackLittermanModel2014a}
\APACinsertmetastar {%
duqiBlackLittermanModel2014a}%
\begin{APACrefauthors}%
Duqi, A.%
, Franci, L.%
\BCBL {} Torluccio, G.%
\end{APACrefauthors}%
\unskip\
\newblock
\APACrefYearMonthDay{2014}{}{}.
\newblock
{\BBOQ}\APACrefatitle {The Black--Litterman model: the definition of views based on volatility forecasts} {The black--litterman model: the definition of views based on volatility forecasts}.{\BBCQ}
\newblock
\APACjournalVolNumPages{Applied Financial Economics}{24}{19}{1285--1296,}
\newblock
\begin{APACrefDOI} \doi{10.1080/09603107.2014.925056} \end{APACrefDOI}
\newblock

\newblock

\PrintBackRefs{\CurrentBib}

\bibitem [\protect \citeauthoryear {%
Gao%
, Zhang%
\BCBL {}\ \BBA {} Yang%
}{%
Gao%
\ \protect \BOthers {.}}{%
{\protect \APACyear {2020}}%
}]{%
gao2020application}
\APACinsertmetastar {%
gao2020application}%
\begin{APACrefauthors}%
Gao, P.%
, Zhang, R.%
\BCBL {} Yang, X.%
\end{APACrefauthors}%
\unskip\
\newblock
\APACrefYearMonthDay{2020}{}{}.
\newblock
{\BBOQ}\APACrefatitle {The application of stock index price prediction with neural network} {The application of stock index price prediction with neural network}.{\BBCQ}
\newblock
\APACjournalVolNumPages{Mathematical and Computational Applications}{25}{3}{53,}
\newblock

\newblock

\PrintBackRefs{\CurrentBib}

\bibitem [\protect \citeauthoryear {%
Hassani%
, Dionisio%
\BCBL {}\ \BBA {} Ghodsi%
}{%
Hassani%
\ \protect \BOthers {.}}{%
{\protect \APACyear {2010}}%
}]{%
hassaniEffectNoiseReduction2010}
\APACinsertmetastar {%
hassaniEffectNoiseReduction2010}%
\begin{APACrefauthors}%
Hassani, H.%
, Dionisio, A.%
\BCBL {} Ghodsi, M.%
\end{APACrefauthors}%
\unskip\
\newblock
\APACrefYearMonthDay{2010}{}{}.
\newblock
{\BBOQ}\APACrefatitle {The effect of noise reduction in measuring the linear and nonlinear dependency of financial markets} {The effect of noise reduction in measuring the linear and nonlinear dependency of financial markets}.{\BBCQ}
\newblock
\APACjournalVolNumPages{Nonlinear Analysis: Real World Applications}{11}{1}{492--502,}
\newblock
\begin{APACrefDOI} \doi{10.1016/j.nonrwa.2009.01.004} \end{APACrefDOI}
\newblock

\newblock

\PrintBackRefs{\CurrentBib}

\bibitem [\protect \citeauthoryear {%
He%
\ \BBA {} Litterman%
}{%
He%
\ \BBA {} Litterman%
}{%
{\protect \APACyear {2002}}%
}]{%
heIntuitionBlackLittermanModel2002}
\APACinsertmetastar {%
heIntuitionBlackLittermanModel2002}%
\begin{APACrefauthors}%
He, G.%
\BCBT {}\ \BBA {} Litterman, R.%
\end{APACrefauthors}%
\unskip\
\newblock
\APACrefYearMonthDay{2002}{}{}.
\newblock
{\BBOQ}\APACrefatitle {The intuition behind Black-Litterman model portfolios} {The intuition behind black-litterman model portfolios}.{\BBCQ}
\newblock
\APACjournalVolNumPages{Available at SSRN 334304}{}{}{,}
\newblock
\begin{APACrefDOI} \doi{10.2139/ssrn.334304} \end{APACrefDOI}
\newblock

\newblock

\PrintBackRefs{\CurrentBib}

\bibitem [\protect \citeauthoryear {%
Huang%
\ \protect \BOthers {.}}{%
Huang%
\ \protect \BOthers {.}}{%
{\protect \APACyear {1998}}%
}]{%
huangEmpiricalModeDecomposition1998}
\APACinsertmetastar {%
huangEmpiricalModeDecomposition1998}%
\begin{APACrefauthors}%
Huang, N.E.%
, Shen, Z.%
, Long, S.R.%
, Wu, M.C.%
, Shih, H.H.%
, Zheng, Q.%
\BDBL {}Liu, H.H.%
\end{APACrefauthors}%
\unskip\
\newblock
\APACrefYearMonthDay{1998}{}{}.
\newblock
{\BBOQ}\APACrefatitle {The empirical mode decomposition and the Hilbert spectrum for nonlinear and non-stationary time series analysis} {The empirical mode decomposition and the hilbert spectrum for nonlinear and non-stationary time series analysis}.{\BBCQ}
\newblock
\APACjournalVolNumPages{Proceedings of the Royal Society of London. Series A: mathematical, physical and engineering sciences}{454}{1971}{903--995,}
\newblock
\begin{APACrefDOI} \doi{10.1098/rspa.1998.0193} \end{APACrefDOI}
\newblock

\newblock

\PrintBackRefs{\CurrentBib}

\bibitem [\protect \citeauthoryear {%
Idzorek%
}{%
Idzorek%
}{%
{\protect \APACyear {2007}}%
}]{%
idzorek2StepbystepGuide2007}
\APACinsertmetastar {%
idzorek2StepbystepGuide2007}%
\begin{APACrefauthors}%
Idzorek, T.%
\end{APACrefauthors}%
\unskip\
\newblock
\APACrefYearMonthDay{2007}{}{}.
\newblock
{\BBOQ}\APACrefatitle {A step-by-step guide to the Black-Litterman model: Incorporating user-specified confidence levels} {A step-by-step guide to the black-litterman model: Incorporating user-specified confidence levels}.{\BBCQ}
\newblock
 \APACrefbtitle {Forecasting expected returns in the financial markets} {Forecasting expected returns in the financial markets}\ (\BPGS\ 17--38).
\newblock
\APACaddressPublisher{}{Elsevier}.
\PrintBackRefs{\CurrentBib}

\bibitem [\protect \citeauthoryear {%
Jiang%
}{%
Jiang%
}{%
{\protect \APACyear {2021}}%
}]{%
jiangApplicationsDeepLearning2021}
\APACinsertmetastar {%
jiangApplicationsDeepLearning2021}%
\begin{APACrefauthors}%
Jiang, W.%
\end{APACrefauthors}%
\unskip\
\newblock
\APACrefYearMonthDay{2021}{}{}.
\newblock
{\BBOQ}\APACrefatitle {Applications of deep learning in stock market prediction: recent progress} {Applications of deep learning in stock market prediction: recent progress}.{\BBCQ}
\newblock
\APACjournalVolNumPages{Expert Systems with Applications}{184}{}{115537,}
\newblock
\begin{APACrefDOI} \doi{10.1016/j.eswa.2021.115537} \end{APACrefDOI}
\newblock

\newblock

\PrintBackRefs{\CurrentBib}

\bibitem [\protect \citeauthoryear {%
Kara%
, Ulucan%
\BCBL {}\ \BBA {} Atici%
}{%
Kara%
\ \protect \BOthers {.}}{%
{\protect \APACyear {2019}}%
}]{%
karaHybridApproachGenerating2019}
\APACinsertmetastar {%
karaHybridApproachGenerating2019}%
\begin{APACrefauthors}%
Kara, M.%
, Ulucan, A.%
\BCBL {} Atici, K.B.%
\end{APACrefauthors}%
\unskip\
\newblock
\APACrefYearMonthDay{2019}{}{}.
\newblock
{\BBOQ}\APACrefatitle {A hybrid approach for generating investor views in Black--Litterman model} {A hybrid approach for generating investor views in black--litterman model}.{\BBCQ}
\newblock
\APACjournalVolNumPages{Expert Systems with Applications}{128}{}{256--270,}
\newblock
\begin{APACrefDOI} \doi{10.1016/j.eswa.2019.03.041} \end{APACrefDOI}
\newblock

\newblock

\PrintBackRefs{\CurrentBib}

\bibitem [\protect \citeauthoryear {%
Kumbure%
, Lohrmann%
, Luukka%
\BCBL {}\ \BBA {} Porras%
}{%
Kumbure%
\ \protect \BOthers {.}}{%
{\protect \APACyear {2022}}%
}]{%
kumbureMachineLearningTechniques2022}
\APACinsertmetastar {%
kumbureMachineLearningTechniques2022}%
\begin{APACrefauthors}%
Kumbure, M.M.%
, Lohrmann, C.%
, Luukka, P.%
\BCBL {} Porras, J.%
\end{APACrefauthors}%
\unskip\
\newblock
\APACrefYearMonthDay{2022}{}{}.
\newblock
{\BBOQ}\APACrefatitle {Machine learning techniques and data for stock market forecasting: A literature review} {Machine learning techniques and data for stock market forecasting: A literature review}.{\BBCQ}
\newblock
\APACjournalVolNumPages{Expert Systems with Applications}{197}{}{116659,}
\newblock
\begin{APACrefDOI} \doi{10.1016/j.eswa.2022.116659} \end{APACrefDOI}
\newblock

\newblock

\PrintBackRefs{\CurrentBib}

\bibitem [\protect \citeauthoryear {%
Lahmiri%
}{%
Lahmiri%
}{%
{\protect \APACyear {2018}}%
}]{%
lahmiriMinuteaheadStockPrice2018}
\APACinsertmetastar {%
lahmiriMinuteaheadStockPrice2018}%
\begin{APACrefauthors}%
Lahmiri, S.%
\end{APACrefauthors}%
\unskip\
\newblock
\APACrefYearMonthDay{2018}{}{}.
\newblock
{\BBOQ}\APACrefatitle {Minute-ahead stock price forecasting based on singular spectrum analysis and support vector regression} {Minute-ahead stock price forecasting based on singular spectrum analysis and support vector regression}.{\BBCQ}
\newblock
\APACjournalVolNumPages{Applied Mathematics and Computation}{320}{}{444--451,}
\newblock
\begin{APACrefDOI} \doi{10.1016/j.amc.2017.09.049} \end{APACrefDOI}
\newblock

\newblock

\PrintBackRefs{\CurrentBib}

\bibitem [\protect \citeauthoryear {%
Lang%
\ \protect \BOthers {.}}{%
Lang%
\ \protect \BOthers {.}}{%
{\protect \APACyear {2018}}%
}]{%
langFastMultivariateEmpirical2018}
\APACinsertmetastar {%
langFastMultivariateEmpirical2018}%
\begin{APACrefauthors}%
Lang, X.%
, Zheng, Q.%
, Zhang, Z.%
, Lu, S.%
, Xie, L.%
, Horch, A.%
\BCBL {} Su, H.%
\end{APACrefauthors}%
\unskip\
\newblock
\APACrefYearMonthDay{2018}{}{}.
\newblock
{\BBOQ}\APACrefatitle {Fast multivariate empirical mode decomposition} {Fast multivariate empirical mode decomposition}.{\BBCQ}
\newblock
\APACjournalVolNumPages{IEEE Access}{6}{}{65521--65538,}
\newblock
\begin{APACrefDOI} \doi{10.1109/ACCESS.2018.2877150} \end{APACrefDOI}
\newblock

\newblock

\PrintBackRefs{\CurrentBib}

\bibitem [\protect \citeauthoryear {%
Lei%
}{%
Lei%
}{%
{\protect \APACyear {2019}}%
}]{%
leiBlackLittermanAsset2019}
\APACinsertmetastar {%
leiBlackLittermanAsset2019}%
\begin{APACrefauthors}%
Lei, D.%
\end{APACrefauthors}%
\unskip\
\newblock
\APACrefYearMonthDay{2019}{}{}.
\newblock
{\BBOQ}\APACrefatitle {Black--Litterman asset allocation model based on principal component analysis (PCA) under uncertainty} {Black--litterman asset allocation model based on principal component analysis (pca) under uncertainty}.{\BBCQ}
\newblock
\APACjournalVolNumPages{Cluster Computing}{22}{}{4299--4306,}
\newblock
\begin{APACrefDOI} \doi{10.1007/s10586-018-1864-1} \end{APACrefDOI}
\newblock

\newblock

\PrintBackRefs{\CurrentBib}

\bibitem [\protect \citeauthoryear {%
Lin%
, Yan%
, Xu%
, Liao%
\BCBL {}\ \BBA {} Ma%
}{%
Lin%
\ \protect \BOthers {.}}{%
{\protect \APACyear {2021}}%
}]{%
linForecastingStockIndex2021}
\APACinsertmetastar {%
linForecastingStockIndex2021}%
\begin{APACrefauthors}%
Lin, Y.%
, Yan, Y.%
, Xu, J.%
, Liao, Y.%
\BCBL {} Ma, F.%
\end{APACrefauthors}%
\unskip\
\newblock
\APACrefYearMonthDay{2021}{}{}.
\newblock
{\BBOQ}\APACrefatitle {Forecasting stock index price using the CEEMDAN-LSTM model} {Forecasting stock index price using the ceemdan-lstm model}.{\BBCQ}
\newblock
\APACjournalVolNumPages{The North American Journal of Economics and Finance}{57}{}{101421,}
\newblock
\begin{APACrefDOI} \doi{10.1016/j.najef.2021.101421} \end{APACrefDOI}
\newblock

\newblock

\PrintBackRefs{\CurrentBib}

\bibitem [\protect \citeauthoryear {%
Markowitz%
}{%
Markowitz%
}{%
{\protect \APACyear {1952}}%
}]{%
markowitzPortfolioSelection1952}
\APACinsertmetastar {%
markowitzPortfolioSelection1952}%
\begin{APACrefauthors}%
Markowitz, H.%
\end{APACrefauthors}%
\unskip\
\newblock
\APACrefYearMonthDay{1952}{}{}.
\newblock
{\BBOQ}\APACrefatitle {Portfolio Selection} {Portfolio selection}.{\BBCQ}
\newblock
\APACjournalVolNumPages{The Journal of Finance}{7}{1}{77--91,}
\newblock
\begin{APACrefDOI} \doi{10.2307/2975974} \end{APACrefDOI}
\newblock

\newblock

\PrintBackRefs{\CurrentBib}

\bibitem [\protect \citeauthoryear {%
Michaud%
}{%
Michaud%
}{%
{\protect \APACyear {1989}}%
}]{%
michaud1989markowitz}
\APACinsertmetastar {%
michaud1989markowitz}%
\begin{APACrefauthors}%
Michaud, R.O.%
\end{APACrefauthors}%
\unskip\
\newblock
\APACrefYearMonthDay{1989}{}{}.
\newblock
{\BBOQ}\APACrefatitle {The Markowitz optimization enigma: Is ‘optimized’optimal?} {The markowitz optimization enigma: Is ‘optimized’optimal?}{\BBCQ}
\newblock
\APACjournalVolNumPages{Financial analysts journal}{45}{1}{31--42,}
\newblock
\begin{APACrefDOI} \doi{10.2139/ssrn.2387669} \end{APACrefDOI}
\newblock

\newblock

\PrintBackRefs{\CurrentBib}

\bibitem [\protect \citeauthoryear {%
Mourad%
}{%
Mourad%
}{%
{\protect \APACyear {2023}}%
}]{%
mouradMultivariateGroupsparseMode2023}
\APACinsertmetastar {%
mouradMultivariateGroupsparseMode2023}%
\begin{APACrefauthors}%
Mourad, N.%
\end{APACrefauthors}%
\unskip\
\newblock
\APACrefYearMonthDay{2023}{}{}.
\newblock
{\BBOQ}\APACrefatitle {Multivariate group-sparse mode decomposition} {Multivariate group-sparse mode decomposition}.{\BBCQ}
\newblock
\APACjournalVolNumPages{Digital Signal Processing}{137}{}{104024,}
\newblock
\begin{APACrefDOI} \doi{10.1016/j.dsp.2023.104024} \end{APACrefDOI}
\newblock

\newblock

\PrintBackRefs{\CurrentBib}

\bibitem [\protect \citeauthoryear {%
Palomba%
}{%
Palomba%
}{%
{\protect \APACyear {2008}}%
}]{%
palombaMultivariateGARCHModels2008}
\APACinsertmetastar {%
palombaMultivariateGARCHModels2008}%
\begin{APACrefauthors}%
Palomba, G.%
\end{APACrefauthors}%
\unskip\
\newblock
\APACrefYearMonthDay{2008}{}{}.
\newblock
{\BBOQ}\APACrefatitle {Multivariate GARCH models and the Black-Litterman approach for tracking error constrained portfolios: an empirical analysis} {Multivariate garch models and the black-litterman approach for tracking error constrained portfolios: an empirical analysis}.{\BBCQ}
\newblock
\APACjournalVolNumPages{Global Business and Economics Review}{10}{4}{379--413,}
\newblock
\begin{APACrefDOI} \doi{10.1504/GBER.2008.020592} \end{APACrefDOI}
\newblock

\newblock

\PrintBackRefs{\CurrentBib}

\bibitem [\protect \citeauthoryear {%
Pyo%
\ \BBA {} Lee%
}{%
Pyo%
\ \BBA {} Lee%
}{%
{\protect \APACyear {2018}}%
}]{%
pyoExploitingLowriskAnomaly2018}
\APACinsertmetastar {%
pyoExploitingLowriskAnomaly2018}%
\begin{APACrefauthors}%
Pyo, S.%
\BCBT {}\ \BBA {} Lee, J.%
\end{APACrefauthors}%
\unskip\
\newblock
\APACrefYearMonthDay{2018}{}{}.
\newblock
{\BBOQ}\APACrefatitle {Exploiting the low-risk anomaly using machine learning to enhance the Black--Litterman framework: Evidence from South Korea} {Exploiting the low-risk anomaly using machine learning to enhance the black--litterman framework: Evidence from south korea}.{\BBCQ}
\newblock
\APACjournalVolNumPages{Pacific-Basin Finance Journal}{51}{}{1--12,}
\newblock
\begin{APACrefDOI} \doi{10.1016/j.pacfin.2018.06.002} \end{APACrefDOI}
\newblock

\newblock

\PrintBackRefs{\CurrentBib}

\bibitem [\protect \citeauthoryear {%
Rehman%
\ \BBA {} Mandic%
}{%
Rehman%
\ \BBA {} Mandic%
}{%
{\protect \APACyear {2010}}%
}]{%
rehmanMultivariateEmpiricalMode2010}
\APACinsertmetastar {%
rehmanMultivariateEmpiricalMode2010}%
\begin{APACrefauthors}%
Rehman, N.%
\BCBT {}\ \BBA {} Mandic, D.P.%
\end{APACrefauthors}%
\unskip\
\newblock
\APACrefYearMonthDay{2010}{}{}.
\newblock
{\BBOQ}\APACrefatitle {Multivariate empirical mode decomposition} {Multivariate empirical mode decomposition}.{\BBCQ}
\newblock
\APACjournalVolNumPages{Proceedings of the Royal Society A: Mathematical, Physical and Engineering Sciences}{466}{2117}{1291--1302,}
\newblock
\begin{APACrefDOI} \doi{10.1098/rspa.2009.0502} \end{APACrefDOI}
\newblock

\newblock

\PrintBackRefs{\CurrentBib}

\bibitem [\protect \citeauthoryear {%
Rezaei%
, Faaljou%
\BCBL {}\ \BBA {} Mansourfar%
}{%
Rezaei%
\ \protect \BOthers {.}}{%
{\protect \APACyear {2021}}%
}]{%
rezaeiIntelligentAssetAllocation2021}
\APACinsertmetastar {%
rezaeiIntelligentAssetAllocation2021}%
\begin{APACrefauthors}%
Rezaei, H.%
, Faaljou, H.%
\BCBL {} Mansourfar, G.%
\end{APACrefauthors}%
\unskip\
\newblock
\APACrefYearMonthDay{2021}{}{}.
\newblock
{\BBOQ}\APACrefatitle {Intelligent asset allocation using predictions of deep frequency decomposition} {Intelligent asset allocation using predictions of deep frequency decomposition}.{\BBCQ}
\newblock
\APACjournalVolNumPages{Expert Systems with Applications}{186}{}{115715,}
\newblock
\begin{APACrefDOI} \doi{10.1016/j.eswa.2021.115715} \end{APACrefDOI}
\newblock

\newblock

\PrintBackRefs{\CurrentBib}

\bibitem [\protect \citeauthoryear {%
Rilling%
, Flandrin%
, Gon{\c{c}}alves%
\BCBL {}\ \BBA {} Lilly%
}{%
Rilling%
\ \protect \BOthers {.}}{%
{\protect \APACyear {2007}}%
}]{%
rillingBivariateEmpiricalMode2007}
\APACinsertmetastar {%
rillingBivariateEmpiricalMode2007}%
\begin{APACrefauthors}%
Rilling, G.%
, Flandrin, P.%
, Gon{\c{c}}alves, P.%
\BCBL {} Lilly, J.M.%
\end{APACrefauthors}%
\unskip\
\newblock
\APACrefYearMonthDay{2007}{}{}.
\newblock
{\BBOQ}\APACrefatitle {Bivariate empirical mode decomposition} {Bivariate empirical mode decomposition}.{\BBCQ}
\newblock
\APACjournalVolNumPages{IEEE signal processing letters}{14}{12}{936--939,}
\newblock
\begin{APACrefDOI} \doi{10.1109/LSP.2007.904710} \end{APACrefDOI}
\newblock

\newblock

\PrintBackRefs{\CurrentBib}

\bibitem [\protect \citeauthoryear {%
Satchell%
\ \BBA {} Scowcroft%
}{%
Satchell%
\ \BBA {} Scowcroft%
}{%
{\protect \APACyear {2007}}%
}]{%
satchellDemystificationBlackLitterman2000}
\APACinsertmetastar {%
satchellDemystificationBlackLitterman2000}%
\begin{APACrefauthors}%
Satchell, S.%
\BCBT {}\ \BBA {} Scowcroft, A.%
\end{APACrefauthors}%
\unskip\
\newblock
\APACrefYearMonthDay{2007}{}{}.
\newblock
{\BBOQ}\APACrefatitle {A demystification of the Black-Litterman model: Managing quantitative and traditional portfolio construction} {A demystification of the black-litterman model: Managing quantitative and traditional portfolio construction}.{\BBCQ}
\newblock
 \APACrefbtitle {Forecasting expected returns in the financial markets} {Forecasting expected returns in the financial markets}\ (\BPGS\ 39--53).
\newblock
\APACaddressPublisher{}{Elsevier}.
\PrintBackRefs{\CurrentBib}

\bibitem [\protect \citeauthoryear {%
Shu%
\ \BBA {} Gao%
}{%
Shu%
\ \BBA {} Gao%
}{%
{\protect \APACyear {2020}}%
}]{%
shuForecastingStockPrice2020}
\APACinsertmetastar {%
shuForecastingStockPrice2020}%
\begin{APACrefauthors}%
Shu, W.%
\BCBT {}\ \BBA {} Gao, Q.%
\end{APACrefauthors}%
\unskip\
\newblock
\APACrefYearMonthDay{2020}{}{}.
\newblock
{\BBOQ}\APACrefatitle {Forecasting stock price based on frequency components by EMD and neural networks} {Forecasting stock price based on frequency components by emd and neural networks}.{\BBCQ}
\newblock
\APACjournalVolNumPages{IEEE Access}{8}{}{206388--206395,}
\newblock
\begin{APACrefDOI} \doi{10.1109/ACCESS.2020.3037681} \end{APACrefDOI}
\newblock

\newblock

\PrintBackRefs{\CurrentBib}

\bibitem [\protect \citeauthoryear {%
Q.~Tang%
, Fan%
, Shi%
, Huang%
\BCBL {}\ \BBA {} Ma%
}{%
Q.~Tang%
\ \protect \BOthers {.}}{%
{\protect \APACyear {2021}}%
}]{%
tangPredictionFinancialTime2021}
\APACinsertmetastar {%
tangPredictionFinancialTime2021}%
\begin{APACrefauthors}%
Tang, Q.%
, Fan, T.%
, Shi, R.%
, Huang, J.%
\BCBL {} Ma, Y.%
\end{APACrefauthors}%
\unskip\
\newblock
\APACrefYearMonthDay{2021}{}{}.
\newblock
\APACrefbtitle {Prediction of financial time series using lstm and data denoising methods.} {Prediction of financial time series using lstm and data denoising methods.}
\PrintBackRefs{\CurrentBib}

\bibitem [\protect \citeauthoryear {%
Y.~Tang%
\ \protect \BOthers {.}}{%
Y.~Tang%
\ \protect \BOthers {.}}{%
{\protect \APACyear {2022}}%
}]{%
TANG2022363}
\APACinsertmetastar {%
TANG2022363}%
\begin{APACrefauthors}%
Tang, Y.%
, Song, Z.%
, Zhu, Y.%
, Yuan, H.%
, Hou, M.%
, Ji, J.%
\BDBL {}Li, J.%
\end{APACrefauthors}%
\unskip\
\newblock
\APACrefYearMonthDay{2022}{}{}.
\newblock
{\BBOQ}\APACrefatitle {A survey on machine learning models for financial time series forecasting} {A survey on machine learning models for financial time series forecasting}.{\BBCQ}
\newblock
\APACjournalVolNumPages{Neurocomputing}{512}{}{363-380,}
\newblock
\begin{APACrefDOI} \doi{10.1016/j.neucom.2022.09.003} \end{APACrefDOI}
\newblock

\newblock

\PrintBackRefs{\CurrentBib}

\bibitem [\protect \citeauthoryear {%
Thirumalaisamy%
\ \BBA {} Ansell%
}{%
Thirumalaisamy%
\ \BBA {} Ansell%
}{%
{\protect \APACyear {2018}}%
}]{%
thirumalaisamyFastAdaptiveEmpirical2018}
\APACinsertmetastar {%
thirumalaisamyFastAdaptiveEmpirical2018}%
\begin{APACrefauthors}%
Thirumalaisamy, M.R.%
\BCBT {}\ \BBA {} Ansell, P.J.%
\end{APACrefauthors}%
\unskip\
\newblock
\APACrefYearMonthDay{2018}{}{}.
\newblock
{\BBOQ}\APACrefatitle {Fast and adaptive empirical mode decomposition for multidimensional, multivariate signals} {Fast and adaptive empirical mode decomposition for multidimensional, multivariate signals}.{\BBCQ}
\newblock
\APACjournalVolNumPages{IEEE Signal Processing Letters}{25}{10}{1550--1554,}
\newblock
\begin{APACrefDOI} \doi{10.1109/LSP.2018.2867335} \end{APACrefDOI}
\newblock

\newblock

\PrintBackRefs{\CurrentBib}

\bibitem [\protect \citeauthoryear {%
Torres%
, Colominas%
, Schlotthauer%
\BCBL {}\ \BBA {} Flandrin%
}{%
Torres%
\ \protect \BOthers {.}}{%
{\protect \APACyear {2011}}%
}]{%
torresCompleteEnsembleEmpirical2011a}
\APACinsertmetastar {%
torresCompleteEnsembleEmpirical2011a}%
\begin{APACrefauthors}%
Torres, M.E.%
, Colominas, M.A.%
, Schlotthauer, G.%
\BCBL {} Flandrin, P.%
\end{APACrefauthors}%
\unskip\
\newblock
\APACrefYearMonthDay{2011}{}{}.
\newblock
{\BBOQ}\APACrefatitle {A complete ensemble empirical mode decomposition with adaptive noise} {A complete ensemble empirical mode decomposition with adaptive noise}.{\BBCQ}
\newblock
 \APACrefbtitle {2011 IEEE international conference on acoustics, speech and signal processing (ICASSP)} {2011 ieee international conference on acoustics, speech and signal processing (icassp)}\ (\BPGS\ 4144--4147).
\PrintBackRefs{\CurrentBib}

\bibitem [\protect \citeauthoryear {%
ur Rehman%
\ \BBA {} Mandic%
}{%
ur Rehman%
\ \BBA {} Mandic%
}{%
{\protect \APACyear {2009}}%
}]{%
rehmanEmpiricalModeDecomposition2010}
\APACinsertmetastar {%
rehmanEmpiricalModeDecomposition2010}%
\begin{APACrefauthors}%
ur Rehman, N.%
\BCBT {}\ \BBA {} Mandic, D.P.%
\end{APACrefauthors}%
\unskip\
\newblock
\APACrefYearMonthDay{2009}{}{}.
\newblock
{\BBOQ}\APACrefatitle {Empirical mode decomposition for trivariate signals} {Empirical mode decomposition for trivariate signals}.{\BBCQ}
\newblock
\APACjournalVolNumPages{IEEE Transactions on signal processing}{58}{3}{1059--1068,}
\newblock
\begin{APACrefDOI} \doi{10.1109/TSP.2009.2033730} \end{APACrefDOI}
\newblock

\newblock

\PrintBackRefs{\CurrentBib}

\bibitem [\protect \citeauthoryear {%
C\BHBI H.~Wang%
, Yuan%
, Zeng%
\BCBL {}\ \BBA {} Lin%
}{%
C\BHBI H.~Wang%
\ \protect \BOthers {.}}{%
{\protect \APACyear {2024}}%
}]{%
wangDeepLearningIntegrated2024}
\APACinsertmetastar {%
wangDeepLearningIntegrated2024}%
\begin{APACrefauthors}%
Wang, C\BHBI H.%
, Yuan, J.%
, Zeng, Y.%
\BCBL {} Lin, S.%
\end{APACrefauthors}%
\unskip\
\newblock
\APACrefYearMonthDay{2024}{}{}.
\newblock
{\BBOQ}\APACrefatitle {A deep learning integrated framework for predicting stock index price and fluctuation via singular spectrum analysis and particle swarm optimization} {A deep learning integrated framework for predicting stock index price and fluctuation via singular spectrum analysis and particle swarm optimization}.{\BBCQ}
\newblock
\APACjournalVolNumPages{Applied Intelligence}{54}{2}{1770--1797,}
\newblock
\begin{APACrefDOI} \doi{10.1007/s10489-024-05271-x} \end{APACrefDOI}
\newblock

\newblock

\PrintBackRefs{\CurrentBib}

\bibitem [\protect \citeauthoryear {%
J.~Wang%
\ \BBA {} Liu%
}{%
J.~Wang%
\ \BBA {} Liu%
}{%
{\protect \APACyear {2024}}%
}]{%
wangTwoStageDeepEnsemble2024}
\APACinsertmetastar {%
wangTwoStageDeepEnsemble2024}%
\begin{APACrefauthors}%
Wang, J.%
\BCBT {}\ \BBA {} Liu, J.%
\end{APACrefauthors}%
\unskip\
\newblock
\APACrefYearMonthDay{2024}{}{}.
\newblock
{\BBOQ}\APACrefatitle {Two-stage deep ensemble paradigm based on optimal multi-scale decomposition and multi-factor analysis for stock price prediction} {Two-stage deep ensemble paradigm based on optimal multi-scale decomposition and multi-factor analysis for stock price prediction}.{\BBCQ}
\newblock
\APACjournalVolNumPages{Cognitive Computation}{16}{1}{243--264,}
\newblock
\begin{APACrefDOI} \doi{10.1007/s12559-023-10203-x} \end{APACrefDOI}
\newblock

\newblock

\PrintBackRefs{\CurrentBib}

\bibitem [\protect \citeauthoryear {%
Wu%
\ \BBA {} Huang%
}{%
Wu%
\ \BBA {} Huang%
}{%
{\protect \APACyear {2009}}%
}]{%
wuENSEMBLEEMPIRICALMODE2009}
\APACinsertmetastar {%
wuENSEMBLEEMPIRICALMODE2009}%
\begin{APACrefauthors}%
Wu, Z.%
\BCBT {}\ \BBA {} Huang, N.E.%
\end{APACrefauthors}%
\unskip\
\newblock
\APACrefYearMonthDay{2009}{}{}.
\newblock
{\BBOQ}\APACrefatitle {Ensemble empirical mode decomposition: a noise-assisted data analysis method} {Ensemble empirical mode decomposition: a noise-assisted data analysis method}.{\BBCQ}
\newblock
\APACjournalVolNumPages{Advances in adaptive data analysis}{1}{01}{1--41,}
\newblock
\begin{APACrefDOI} \doi{10.1142/S1793536909000047} \end{APACrefDOI}
\newblock

\newblock

\PrintBackRefs{\CurrentBib}

\bibitem [\protect \citeauthoryear {%
Yao%
, yang Zhang%
\BCBL {}\ \BBA {} Zhao%
}{%
Yao%
\ \protect \BOthers {.}}{%
{\protect \APACyear {2023}}%
}]{%
yaoStockIndexForecasting2023}
\APACinsertmetastar {%
yaoStockIndexForecasting2023}%
\begin{APACrefauthors}%
Yao, Y.%
, yang Zhang, Z.%
\BCBL {} Zhao, Y.%
\end{APACrefauthors}%
\unskip\
\newblock
\APACrefYearMonthDay{2023}{}{}.
\newblock
{\BBOQ}\APACrefatitle {Stock index forecasting based on multivariate empirical mode decomposition and temporal convolutional networks} {Stock index forecasting based on multivariate empirical mode decomposition and temporal convolutional networks}.{\BBCQ}
\newblock
\APACjournalVolNumPages{Applied Soft Computing}{142}{}{110356,}
\newblock
\begin{APACrefDOI} \doi{https://doi.org/10.1016/j.asoc.2023.110356} \end{APACrefDOI}
\newblock

\newblock

\PrintBackRefs{\CurrentBib}

\bibitem [\protect \citeauthoryear {%
Yeh%
, Shieh%
\BCBL {}\ \BBA {} Huang%
}{%
Yeh%
\ \protect \BOthers {.}}{%
{\protect \APACyear {2010}}%
}]{%
yehComplementaryEnsembleEmpirical2010}
\APACinsertmetastar {%
yehComplementaryEnsembleEmpirical2010}%
\begin{APACrefauthors}%
Yeh, J\BHBI R.%
, Shieh, J\BHBI S.%
\BCBL {} Huang, N.E.%
\end{APACrefauthors}%
\unskip\
\newblock
\APACrefYearMonthDay{2010}{}{}.
\newblock
{\BBOQ}\APACrefatitle {Complementary ensemble empirical mode decomposition: A novel noise enhanced data analysis method} {Complementary ensemble empirical mode decomposition: A novel noise enhanced data analysis method}.{\BBCQ}
\newblock
\APACjournalVolNumPages{Advances in adaptive data analysis}{2}{02}{135--156,}
\newblock
\begin{APACrefDOI} \doi{10.1142/S1793536910000422} \end{APACrefDOI}
\newblock

\newblock

\PrintBackRefs{\CurrentBib}

\bibitem [\protect \citeauthoryear {%
Yu%
, Ming%
, Sumei%
\BCBL {}\ \BBA {} Shuping%
}{%
Yu%
\ \protect \BOthers {.}}{%
{\protect \APACyear {2020}}%
}]{%
yuHybridModelFinancial2020}
\APACinsertmetastar {%
yuHybridModelFinancial2020}%
\begin{APACrefauthors}%
Yu, H.%
, Ming, L.J.%
, Sumei, R.%
\BCBL {} Shuping, Z.%
\end{APACrefauthors}%
\unskip\
\newblock
\APACrefYearMonthDay{2020}{}{}.
\newblock
{\BBOQ}\APACrefatitle {A hybrid model for financial time series forecasting—integration of EWT, ARIMA with the improved ABC optimized ELM} {A hybrid model for financial time series forecasting—integration of ewt, arima with the improved abc optimized elm}.{\BBCQ}
\newblock
\APACjournalVolNumPages{IEEE Access}{8}{}{84501--84518,}
\newblock
\begin{APACrefDOI} \doi{10.1109/ACCESS.2020.2987547} \end{APACrefDOI}
\newblock

\newblock

\PrintBackRefs{\CurrentBib}

\bibitem [\protect \citeauthoryear {%
C.~Zhang%
, Sjarif%
\BCBL {}\ \BBA {} Ibrahim%
}{%
C.~Zhang%
\ \protect \BOthers {.}}{%
{\protect \APACyear {2024}}%
}]{%
zhang2024deep}
\APACinsertmetastar {%
zhang2024deep}%
\begin{APACrefauthors}%
Zhang, C.%
, Sjarif, N.N.A.%
\BCBL {} Ibrahim, R.%
\end{APACrefauthors}%
\unskip\
\newblock
\APACrefYearMonthDay{2024}{}{}.
\newblock
{\BBOQ}\APACrefatitle {Deep learning models for price forecasting of financial time series: A review of recent advancements: 2020--2022} {Deep learning models for price forecasting of financial time series: A review of recent advancements: 2020--2022}.{\BBCQ}
\newblock
\APACjournalVolNumPages{Wiley Interdisciplinary Reviews: Data Mining and Knowledge Discovery}{14}{1}{e1519,}
\newblock

\newblock

\PrintBackRefs{\CurrentBib}

\bibitem [\protect \citeauthoryear {%
Y.~Zhang%
, Yan%
\BCBL {}\ \BBA {} Aasma%
}{%
Y.~Zhang%
\ \protect \BOthers {.}}{%
{\protect \APACyear {2020}}%
}]{%
zhangNovelDeepLearning2020}
\APACinsertmetastar {%
zhangNovelDeepLearning2020}%
\begin{APACrefauthors}%
Zhang, Y.%
, Yan, B.%
\BCBL {} Aasma, M.%
\end{APACrefauthors}%
\unskip\
\newblock
\APACrefYearMonthDay{2020}{}{}.
\newblock
{\BBOQ}\APACrefatitle {A novel deep learning framework: Prediction and analysis of financial time series using CEEMD and LSTM} {A novel deep learning framework: Prediction and analysis of financial time series using ceemd and lstm}.{\BBCQ}
\newblock
\APACjournalVolNumPages{Expert systems with applications}{159}{}{113609,}
\newblock
\begin{APACrefDOI} \doi{10.1016/j.eswa.2020.113609} \end{APACrefDOI}
\newblock

\newblock

\PrintBackRefs{\CurrentBib}

\bibitem [\protect \citeauthoryear {%
Zou%
\ \BBA {} He%
}{%
Zou%
\ \BBA {} He%
}{%
{\protect \APACyear {2022}}%
}]{%
zouForecastingCrudeOil2022}
\APACinsertmetastar {%
zouForecastingCrudeOil2022}%
\begin{APACrefauthors}%
Zou, Y.%
\BCBT {}\ \BBA {} He, K.%
\end{APACrefauthors}%
\unskip\
\newblock
\APACrefYearMonthDay{2022}{}{}.
\newblock
{\BBOQ}\APACrefatitle {Forecasting crude oil risk using a multivariate multiscale convolutional neural network model} {Forecasting crude oil risk using a multivariate multiscale convolutional neural network model}.{\BBCQ}
\newblock
\APACjournalVolNumPages{Mathematics}{10}{14}{2413,}
\newblock
\begin{APACrefDOI} \doi{10.3390/math10142413} \end{APACrefDOI}
\newblock

\newblock

\PrintBackRefs{\CurrentBib}

\end{thebibliography}

\end{document}